\documentclass{ws-ijmpa}
\usepackage[super,compress]{cite}
\usepackage{graphicx}

\usepackage{enumerate}
\usepackage[english]{babel}   
\usepackage{xcolor}
\usepackage{colortbl,dcolumn}
\usepackage{textcomp}  
\usepackage{float}
\usepackage{setspace} 
\usepackage{multicol} 
\usepackage{ulem}
\usepackage{bm}    
\usepackage{amsmath,amssymb}
\usepackage{overpic}
\usepackage{subfigure}
\usepackage{multirow}
\usepackage{pifont}
\usepackage{rotating} 
\usepackage{array}
\usepackage{comment}
\usepackage{booktabs}
\usepackage{threeparttable}
\usepackage{wasysym}
\usepackage{lineno} 

\newcommand{\hollowstar}{\text{\ding{73}}}

\graphicspath{{figures/}}
\newsavebox{\tablebox}

\input{alias.tex}

\def\Acp {A_{\CP}}

\def\ATodd {a^{T{\text{-odd}}}_{\CP}}

\def\BR {\mathcal{B}}

\def\Lcp {\Lambda{}_{c}^{+}}
\def\Lcm {\Lbar{}^{-}_{c}}
\def\Xicz {\Xi_{c}^{0}}

\def\Xim {\Xi{}^{-}}

\def\LcToLamKp {{\Lambda_c^+\to\Lambda\Kp}}
\def\LcToLamPip {{\Lambda_c^+\to\Lambda\pip}}
\def\LcToSigKp {{\Lambda_c^+\to\Sigma^{0}\Kp}}
\def\LcToSigPip {{\Lambda_c^+\to\Sigma^{0}\pip}}
\def\LcToLamHp {{\Lambda_c^+\to\Lambda h^+}}
\def\LcToSigHp {{\Lambda_c^+\to\Sigma^{0} h^+}}

\def\LcToLamKp {\Lambda_c^+\to\Lambda\Kp}
\def\LcToLamPip {\Lambda_c^+\to\Lambda\pip}
\def\LcToSigKp {\Lambda_c^+\to\Sigma^{0}\Kp}
\def\LcToSigPip {\Lambda_c^+\to\Sigma^{0}\pip}
\def\LcToLamHp {\Lambda_c^+\to\Lambda h^+}
\def\LcToSigHp {\Lambda_c^+\to\Sigma^{0} h^+}

\begin{document}

\markboth{Longke Li}{Charm physics at the Belle and Belle II experiments}

%
\catchline{}{}{}{}{}
%

\title{Charm physics at the Belle and Belle II experiments\footnote{
presented at the 2024 International Workshop on Future Tau Charm Facilities~(FTCF2024).}}

\author{Longke Li on behalf of the Belle and Belle II Collaborations}

\address{University of Cincinnati \\
Cincinnati, Ohio, 45221-0011, USA\\
lilk@ucmail.uc.edu}

\maketitle

\begin{history}
\received{Day Month Year}
\revised{Day Month Year}
\end{history}

\begin{abstract}
We present recent results on charm physics at the Belle and Belle II experiments, covering measurements of charm lifetimes, branching fractions of the decays of charmed mesons and baryons and the decay asymmetry parameters of two-body decays of charmed baryons, searches for rare and forbidden decays, and measurements of $C\!P$ violating parameters in the four-body decays of charmed mesons and two-body decays of charmed baryons. 

\end{abstract}



\section{Charm production at Belle and Belle II}	
The Belle II experiment, operating in the energy-asymmetric $e^+e^-$ collider SuperKEKB, 
has been designed to conduct precise measurements of weak interaction parameters, explore exotic hadrons, and probe for novel phenomena beyond the Standard Model of particle physics. 
From 2019 to 2022, it accumulated an integrated luminosity 427~$\invfb$,
thereby, a total 1.4~$\invab$ from Belle and Belle II experiments provides large samples of beauty and charm hadrons, as well as tau leptons. 
There exist two primary avenues of charm production at Belle and Belle II: 
(1) via the continuum process $e^+e^-\to{}c\bar{c}$, having a cross section of $\sigma=1.3$~nb; 
(2) from decays of $B$ mesons, where charmed hadrons are involved in the final state. 
In Table~\ref{tab:charmSample}, a comparison of available charm samples at BESIII, Belle, Belle II, and LHCb, along with their own typical characters, is presented. 
Importantly, these experiments will continue to collect data with increased luminosity in the future, heralding a promising outlook for further research in charm physics. 
\begin{table}[htbp]
  \tbl{\label{tab:charmSample}Comparison of available charm samples at BESIII, Belle and Belle II, and LHCb experiments. The typical characters of these three kinds of experiment are also listed.}
{\begin{lrbox}{\tablebox}
\hskip-35pt
\begin{tabular}{ccccccl}  \toprule 
\bf{Experiment}	& \bf{Machine}	& \bf{$E_{C.M.}$}	& \bf{Luminosity}	& \bf{$N_{\rm prod}$} & \bf{Efficiency}		&  \bf{Characters} \\ \midrule
\multirow{4}{*}{
\includegraphics[width=0.13\textwidth]{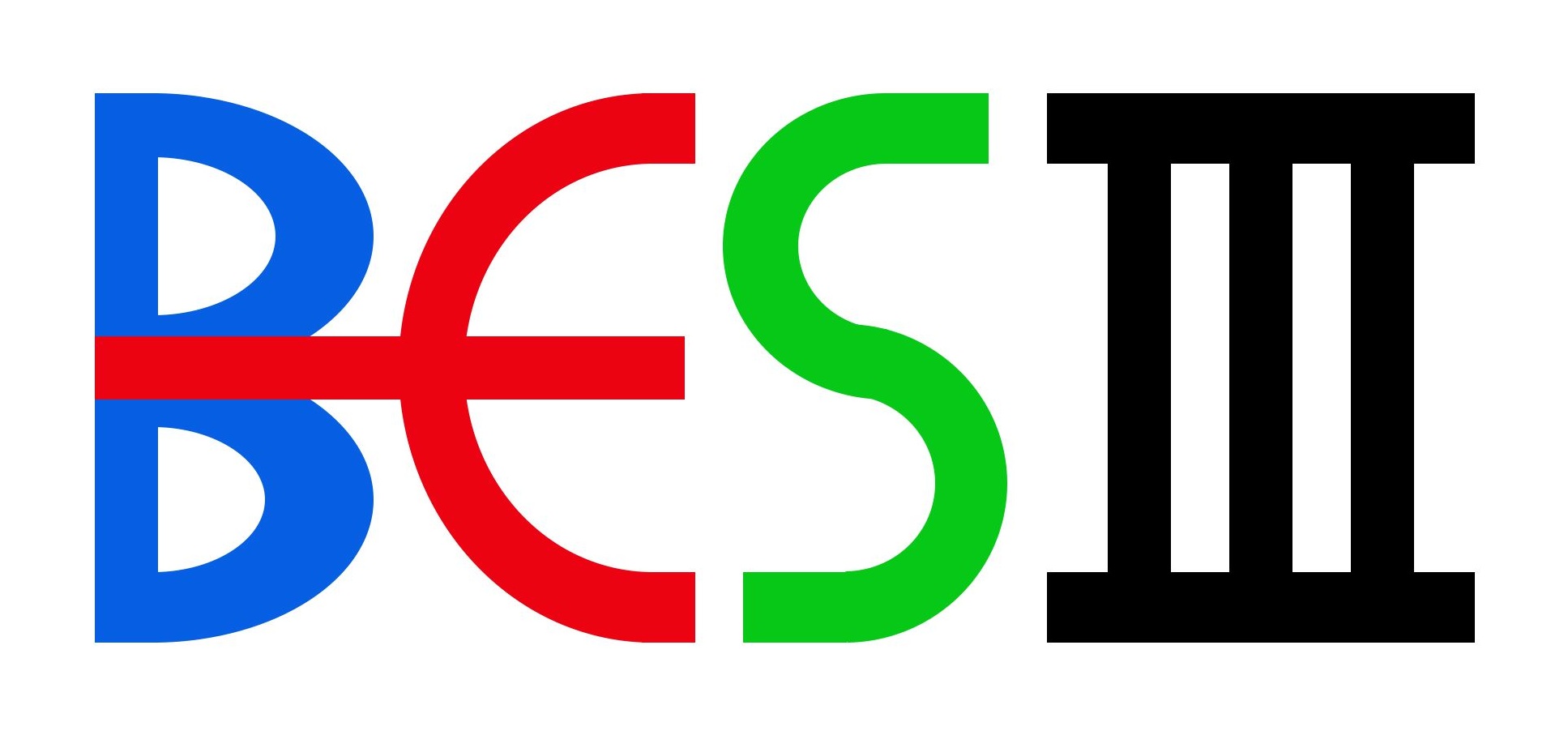}
}			&  	& 3.77 GeV	&  2.9 \textcolor{red}{$(8\to20)$} $\invfb$	& $D^{0,+}$: $10^7$($\to 10^8$)		& \multirow{3}{*}{$\sim10\text{-}30\%$}	&    \textcolor{blue}{$\smiley$} extremely clean environment  \\
			& BEPC-II		&    4.18-4.23 GeV	&  7.3  $\invfb$	&	$D_s^+$: $5\times10^6$ 		&	&  \textcolor{blue}{$\smiley$}  quantum coherence  \\
			& ($\epem$)	&    4.6-4.7 GeV	&  4.5  $\invfb$	&	$\Lcp:$\, $0.8\times10^6$ 		&	&  \textcolor{magenta}{$\frownie$} no boost, no time-dept analysis \\
			&		 	&  		&			&	\textcolor{red}{$\bigstar\hollowstar$}		&	\textcolor{red}{$\bigstar\bigstar\bigstar$}   &  \\ \midrule 	
\multirow{3}{*}{
\includegraphics[width=0.10\textwidth]{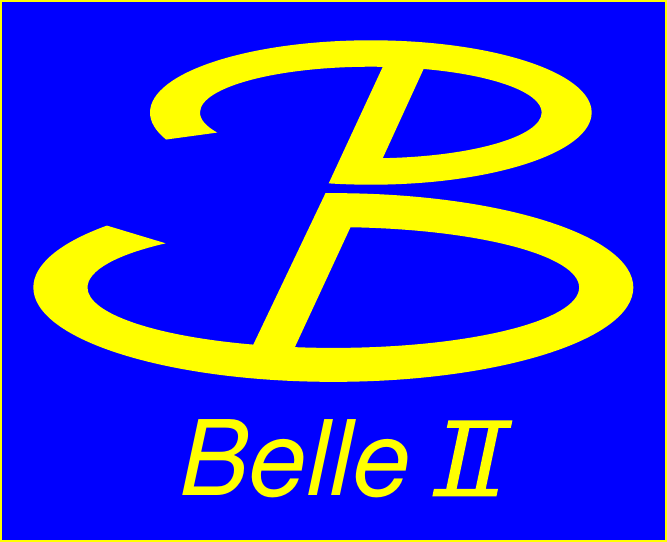}
}			& SuperKEKB	 	&  \multirow{2}{*}{10.58 GeV}	&	\multirow{2}{*}{0.4 \textcolor{red}{($\to50$)} $\invab$}	&	$D^0$: $6\times10^8$ ($\to10^{11}$)	& \multirow{6}{*}{ $\mathcal{O}$(1-10\%) }	& \textcolor{blue}{$\smiley$} high-efficiency detection of neutrals \\
			& ($\epem$) 	&  		&		&		$D_{(s)}^+$: $10^8$ ($\to10^{10}$)	&	 & \textcolor{blue}{$\smiley$} good trigger efficiency \\ 
			& 		 	&    	&  	&	 $\Lambda_c^+$: $10^7$ ($\to10^9$)		&		&   \textcolor{blue}{$\smiley$} time-dependent analysis \\ \cline{2-5}
\multirow{3}{*}{
\includegraphics[width=0.10\textwidth]{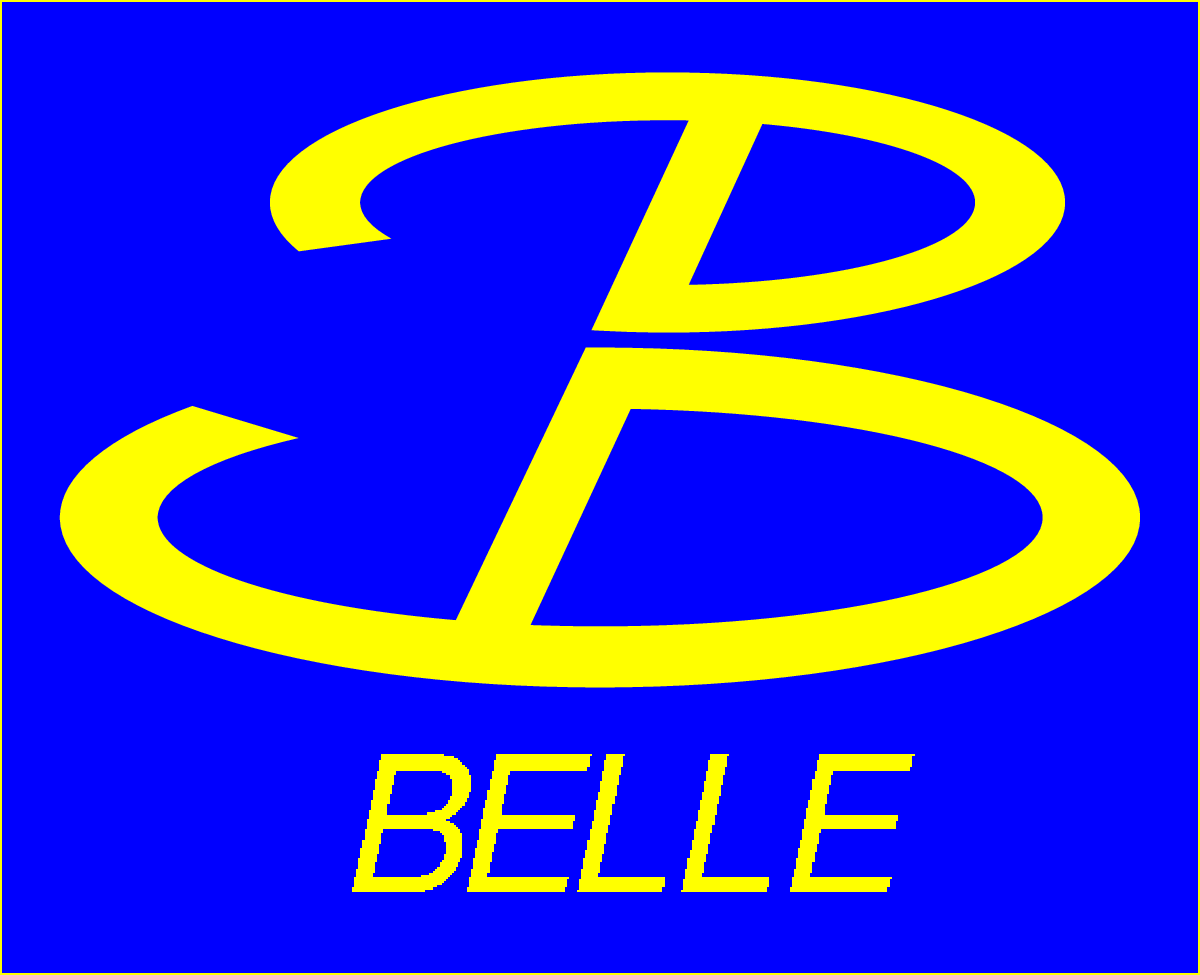}
}			& KEKB	 	&  \multirow{2}{*}{10.58 GeV}	&	\multirow{2}{*}{1  $\invab$}	&	$D^{0,+},D_{s}^+$: $10^9$	&  &  \textcolor{magenta}{$\frownie$} smaller cross-section than LHCb \\
			& ($\epem$) 	&  		&		&	$\Lambda_c^+$: $10^8$	&	& 	 \\ 
			&		 	&  		&		&	\textcolor{red}{$\bigstar\bigstar\hollowstar$}		&  \textcolor{red}{$\bigstar\bigstar$} & 	\\ \midrule	
\multirow{4}{*}{
\includegraphics[width=0.11\textwidth]{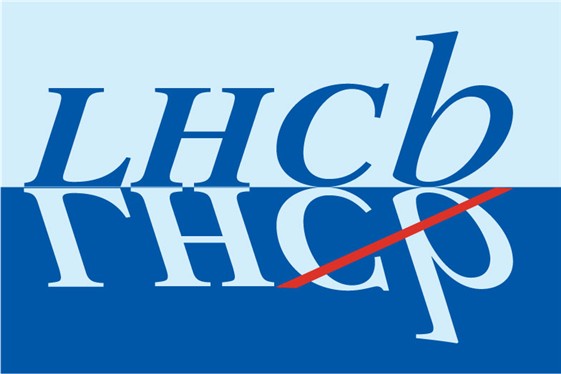}
}			& LHC 		&  7+8 TeV	&	1+2  $\invfb$	&	$5\times10^{12}$		&  \multirow{3}{*}{$\mathcal{O}(0.1\%)$}	&    \textcolor{blue}{$\smiley$} very large production cross-section \\ 	
			& ($pp$) 		&  13 TeV	&	6  $\invfb$	&		$10^{13}$	&	&   \textcolor{blue}{$\smiley$} large boost, excellent time resolution \\ 
			&  			& 	& \textcolor{red}{($\to23\to50$)}~$\invfb$ &			&	& \textcolor{magenta}{$\frownie$} dedicated trigger required \\ 
			&		 	&  		&		&	\textcolor{red}{$\bigstar\bigstar\bigstar\bigstar$}		&	\textcolor{red}{$\bigstar$}   & \\ \bottomrule  
\end{tabular}
\end{lrbox}
\begin{threeparttable}
  \scalebox{0.75}{\usebox{\tablebox}}  
  \begin{tablenotes}
  \begin{spacing}{0.5}
  \begin{footnotesize}
  \item {\tiny{Here uses $\sigma(\Dz\Dzb@3.77\,{\rm GeV})$=3.61~nb, $\sigma(\Dp\Dm@3.77\,{\rm GeV})$=2.88 nb, $\sigma(D_s^{*}D_{s}@4.17\,{\rm GeV})$=0.967~nb; $\sigma(c\bar{c}@10.58\,{\rm GeV})$=1.3~nb where each $c\bar{c}$ event averagely has 1.1/0.6/0.3 $D^0\!/\!D^+\!/\!D_s^+$ yields; $\sigma(D^0@CDF)$=13.3~$\mu$b, and $\sigma(D^0@LHCb)$=1661~$\mu$b, mainly from {\it Int.\,J.\,Mod.\,Phys.\,A\,{\bf 29}(2014)24,14300518}.}}
  \end{footnotesize}
  \end{spacing}
\end{tablenotes}
\end{threeparttable}
}
\end{table}

\section{Charm lifetime measurements}
Hadron lifetimes are difficult to calculate theoretically, as they depend on nonperturbative effects arising from quantum chromodynamics (QCD). 
Comparing calculated and measured values improves our understanding of QCD.
At Belle II, the decay-time resolution is about twice better than that at Belle and BABAR. 
Utilizing the early Belle~II dataset, 
three world-leading charm lifetimes have been measured: $\tau(D^0)=410.5\pm1.1\pm0.8$~fs, 
$\tau(D^+)=1030.4\pm4.7\pm3.1$~fs, and $\tau(\Lcp)=203.20\pm0.89\pm0.77$~fs~\cite{Belle-II:2021cxx,Belle-II:2022ggx}; 
and also a measurement~\cite{Belle-II:2022plj} of $\tau(\Omega_c^0)=410.5\pm1.1\pm0.8$~fs agrees with the measurement by LHCb~\cite{LHCb:2018nfa} and confirm that the $\Omega_c^0$ is not the shortest-lived weakly decaying charmed baryon.

\begin{figure}[!hbtp]  
   \begin{center}
    \begin{overpic}[width=0.33\textwidth,height=0.25\textwidth]{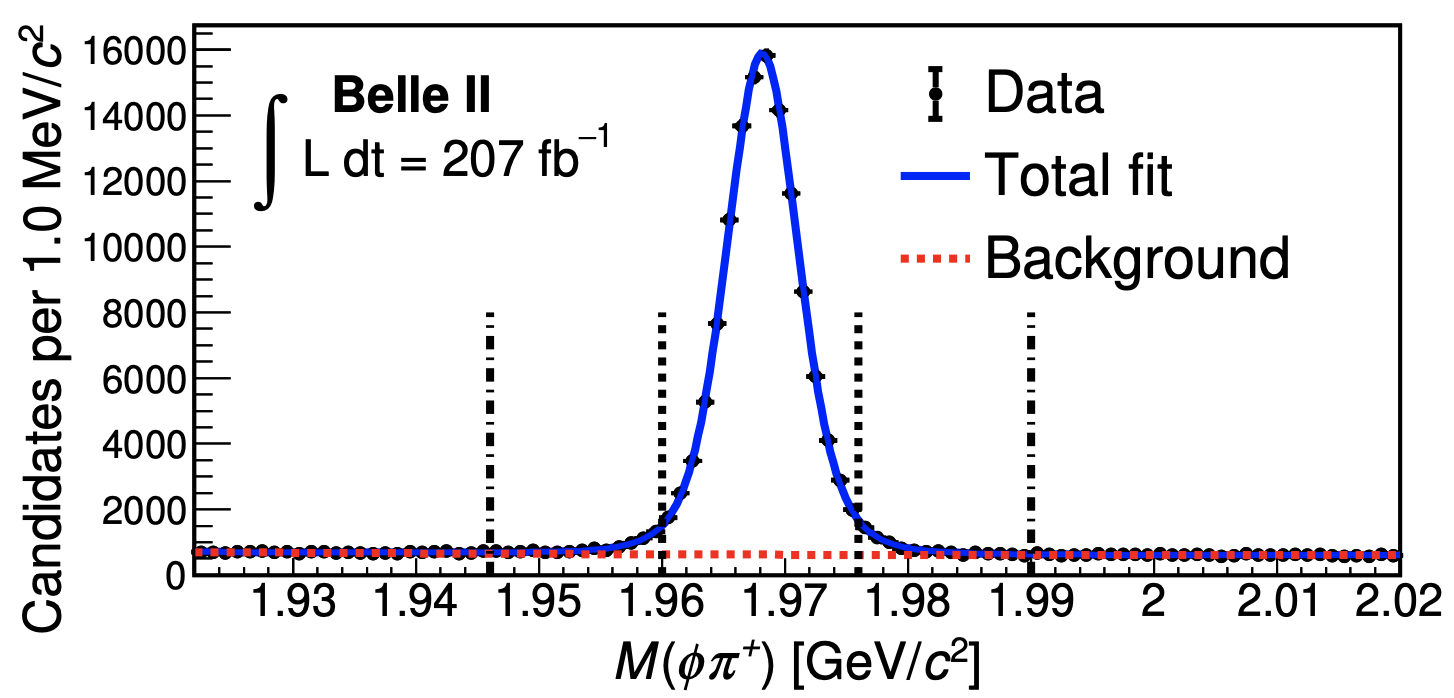}
    \end{overpic}%
    \begin{overpic}[width=0.33\textwidth,height=0.25\textwidth]{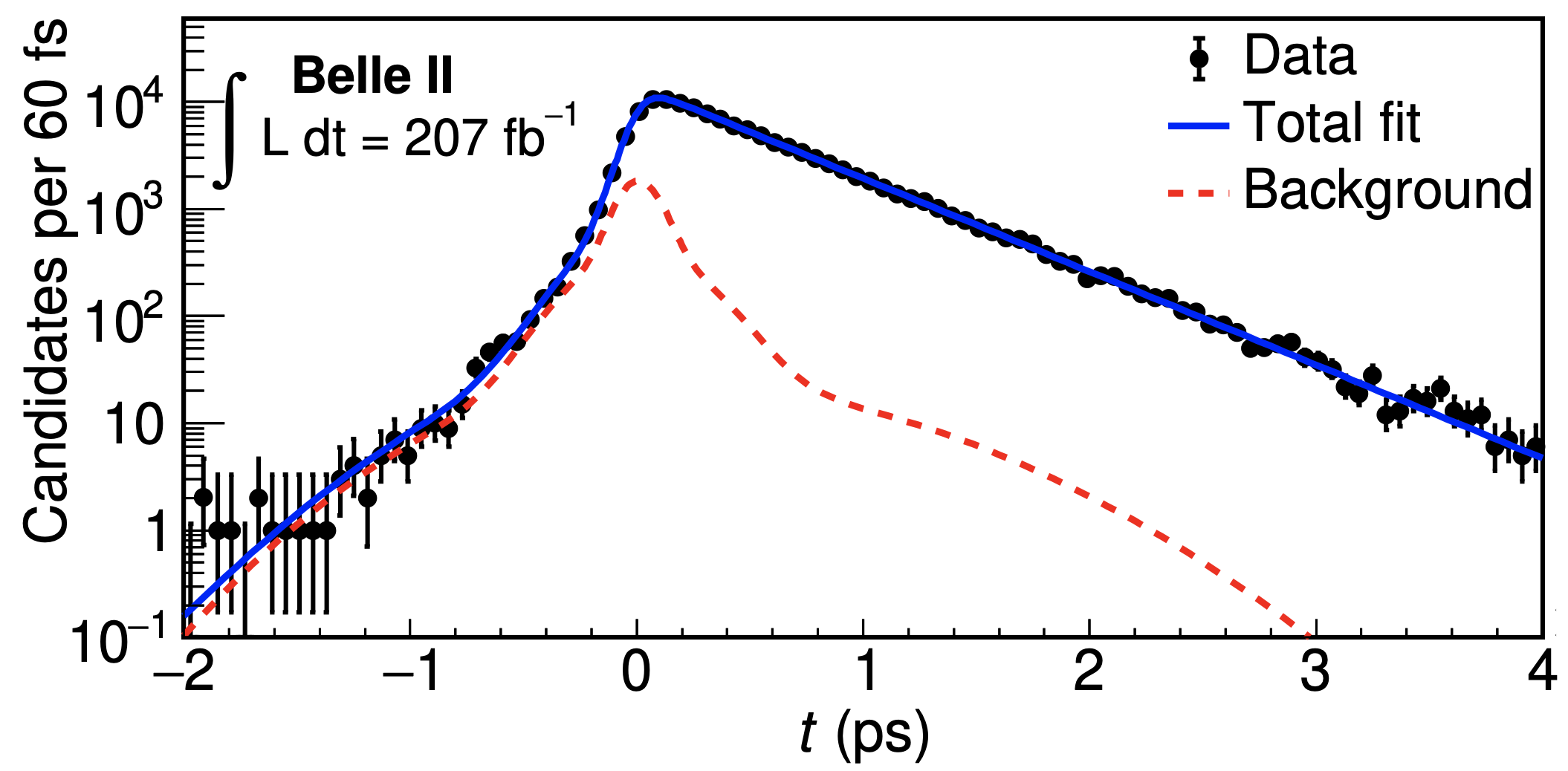}
    \end{overpic}
    \begin{overpic}[width=0.33\textwidth,height=0.25\textwidth]{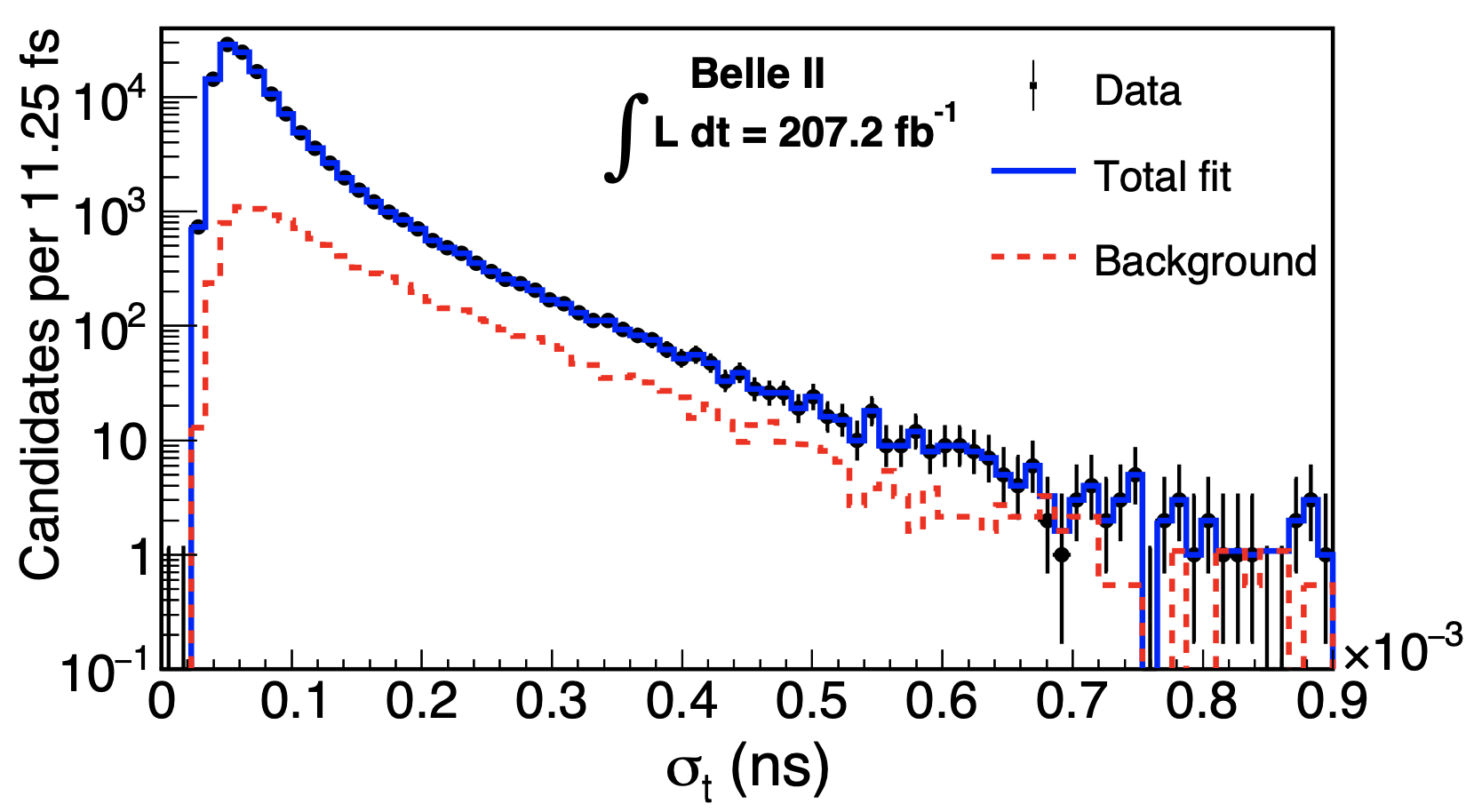}
    \end{overpic}
    \vskip-5pt
    \caption{\label{fig:tauDs}Invariant mass of reconstructed $D_s^+\to\phi\pip$ candidates; the projections of lifetime extraction with a fitting on ($t$, $\sigma_t$) at Belle~II.~\cite{Belle-II:2023eii}}    
    \end{center}
\end{figure}

Based on a clean sample of 116k $D_s^+\to\phi\pip$ reconstructed in 207~$\invfb$ of data at Belle II, the $D_s^+$ lifetime is extracted via an unbinned maximum likelihood fit to the lifetime ($t$) and its uncertainty ($\sigma_t$)~\cite{Belle-II:2023eii}. 
 The likelihood function for $i$th event is calculated by:
  \begin{eqnarray}
  \mathcal{L}(\tau| t^i, \sigma_t^i) = f_{\rm sig} P_{\rm sig}(t^i | \tau, \sigma_t^i) P_{\rm sig}(\sigma_t^i) +  (1-f_{\rm sig}) P_{\rm bkg}(t^i | \tau, \sigma_t^i) P_{\rm bkg}(\sigma_t^i) \nonumber 
  \end{eqnarray}
where $P_{\rm sig}(\sigma_t^i)$ and $P_{\rm bkg}(\sigma_t^i)$ exist to avoid the Punzi bias.
The fitted results are shown in Figure~\ref{fig:tauDs}, and we obtain $\tau_{D_s^+} = (499.5 \pm 1.7 \pm 0.9)$~fs, the world most precise measurement to date.
Thus, Belle~II has made the world's most precise measurements of the $D^{0,+}\,D_s^+,\,\Lambda_c^+$ lifetimes; their small systematic uncertainty demonstrates the excellent performance and understanding of the Belle II detector.

\section{Measurement of branching fraction and decay asymmetry parameter}

\subsection{Branching fraction of Cabibbo-suppressed decays of charmed mesons}
\begin{figure}[!htpb]
\begin{center}
      \begin{overpic}[width=0.4\textwidth,height=0.28\textwidth]{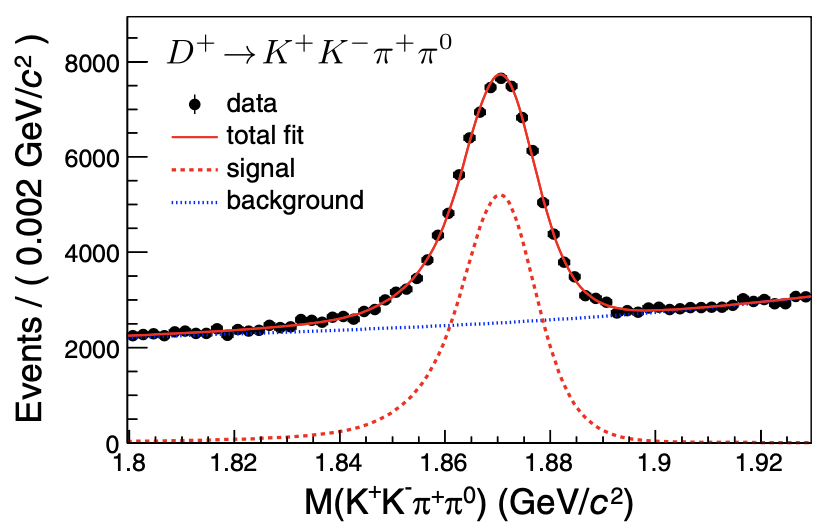}
    \end{overpic}~~%
     \begin{overpic}[width=0.4\textwidth,height=0.28\textwidth]{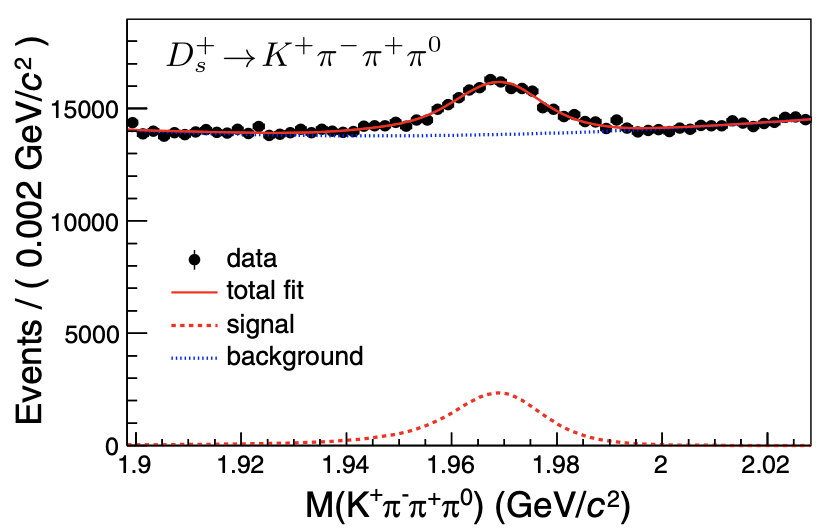}
   \end{overpic}\\
   \vskip2pt
     \begin{overpic}[width=0.4\textwidth,height=0.27\textwidth]{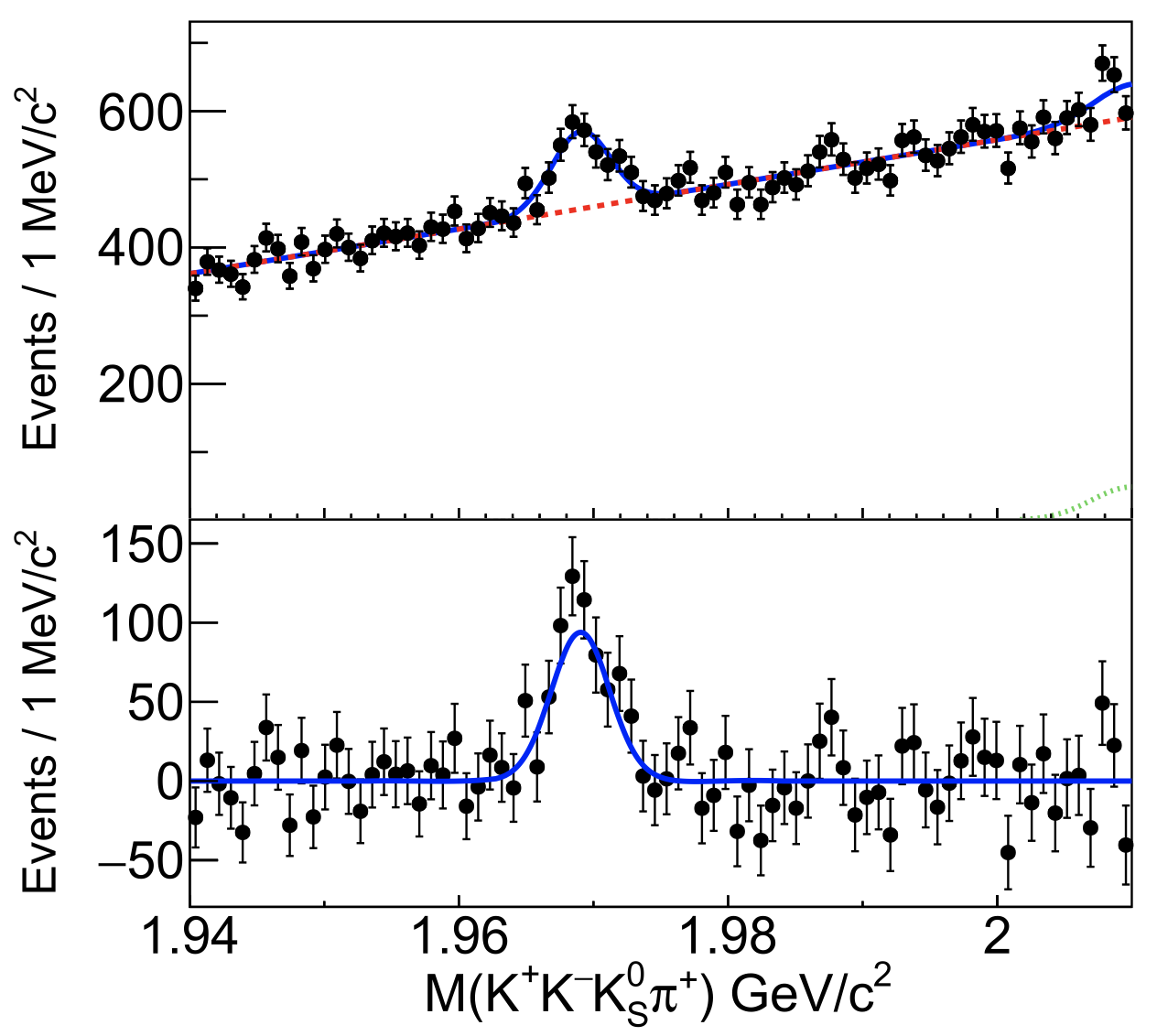}
    \put(18,62){\tiny{$D_s^+\to\KS\Kp\Km\pip$}}
   \end{overpic}~~%
     \begin{overpic}[width=0.4\textwidth,height=0.28\textwidth]{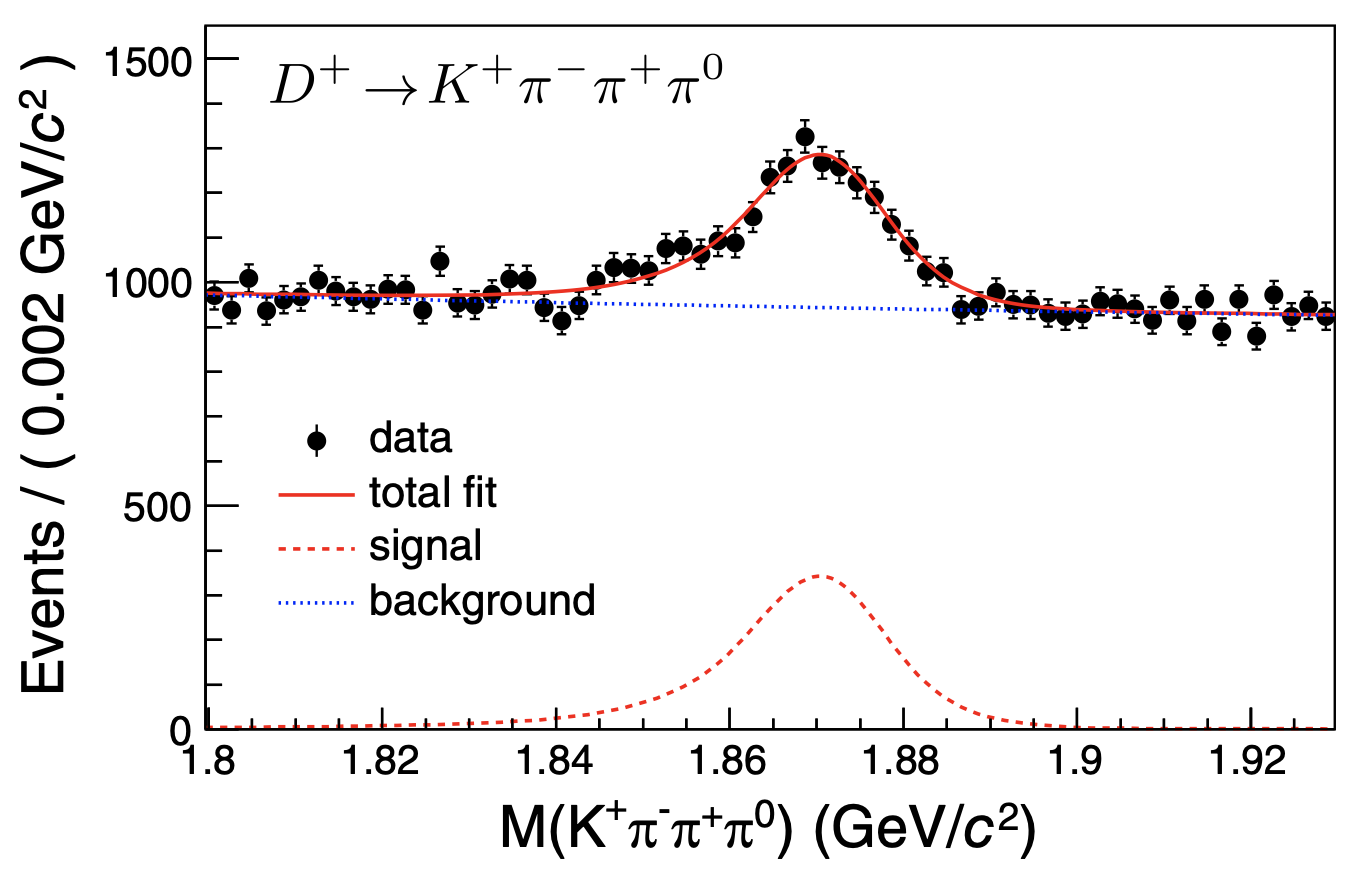}
   \end{overpic}
   \vskip-5pt
\caption{\label{fig:BrCS}Invariant mass of reconstructed $D$ candidates for the SCS decays $D^+\to\Kp\Km\pip\piz$, $D_s^+\to\Kp\pim\pip\piz$, $D_s^+\to\Kp\Km\KS\pip$, and the DCS decay $D^+\to\Kp\pim\pip\piz$ at Belle.~\cite{Belle:2022aha,Belle:2023bzn}}
\end{center}
\end{figure}

Cabibbo-suppressed (CS) hadronic decays of charm mesons offer a potent avenue for exploring new physics. 
Precise measurements of their branching fractions are of paramount importance.
Singly Cabibbo-suppressed (SCS) charm decays serve as essential probes to search for charm $\CP$ violation (CPV) and probe physics beyond the SM.
The abundant charm sample available from Belle and Belle~II provides an excellent opportunity to accurately measure their branching fractions.
Recently, Belle reported several first or most precise branching fractions of charmed meson decays, based on the full dataset.
The invariant mass distributions of reconstructed decays are shown in Figure~\ref{fig:BrCS}.
Using the corresponding well-measured reference modes, we obtain branching fractions ($\BR$) of three SCS decays~\cite{Belle:2022aha,Belle:2023bzn}:
\begin{eqnarray}
\BR(D^+\to\Kp\Km\pip\piz)		& = & (7.08\pm 0.08\pm 0.16\pm 0.20)\times10^{-3}\,, \\
\BR(D_s^+\to\Kp\pim\pip\piz)	& = & (9.44\pm 0.34\pm 0.28\pm 0.32)\times10^{-3}\,,  \\
\BR(D_s^+\to\Kp\Km\KS\pip)	& = & (1.29\pm 0.14\pm 0.04 \pm 0.11)\times 10^{-4}\,;
\end{eqnarray}
and one DCS decays~\cite{Belle:2022aha}: 
\begin{eqnarray}
\BR(D^+\to\Kp\pim\pip\piz)	& = & (1.05\pm 0.07\pm 0.02\pm 0.03)\times10^{-3}\,,
\end{eqnarray}
where the last one confirms the BESIII finding~\cite{BESIII:2020wnc,BESIII:2021uix} of a significantly larger $\BR$ than other known DCS decays.
  
\subsubsection{Branching fraction of charmed baryon decays}
\begin{figure}[!htpb]
\begin{center}
    \begin{overpic}[width=0.33\textwidth,height=0.25\textwidth]{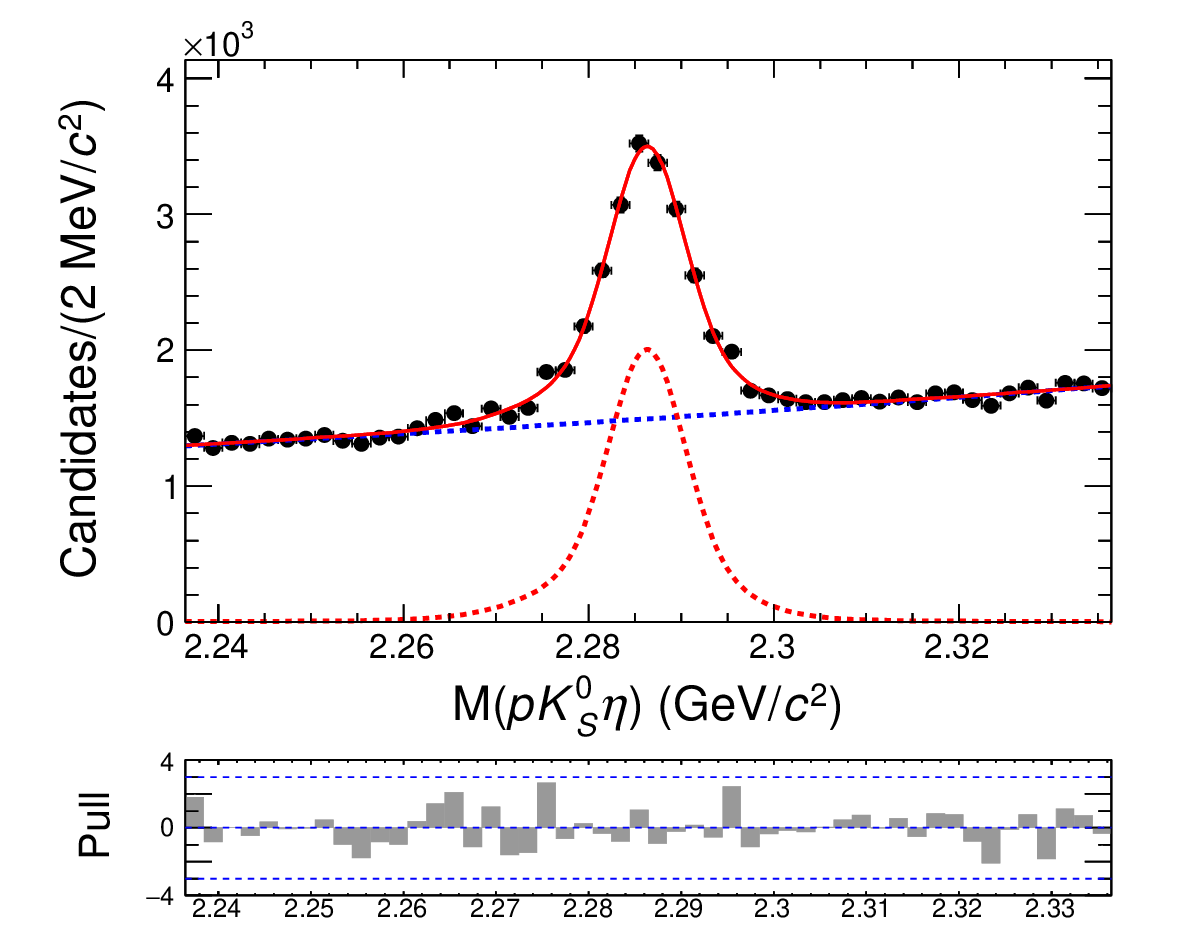}
    \put(18,65){\footnotesize{$\Lcp\to{}p\KS\eta$}}
    \end{overpic}%
     \begin{overpic}[width=0.33\textwidth,height=0.25\textwidth]{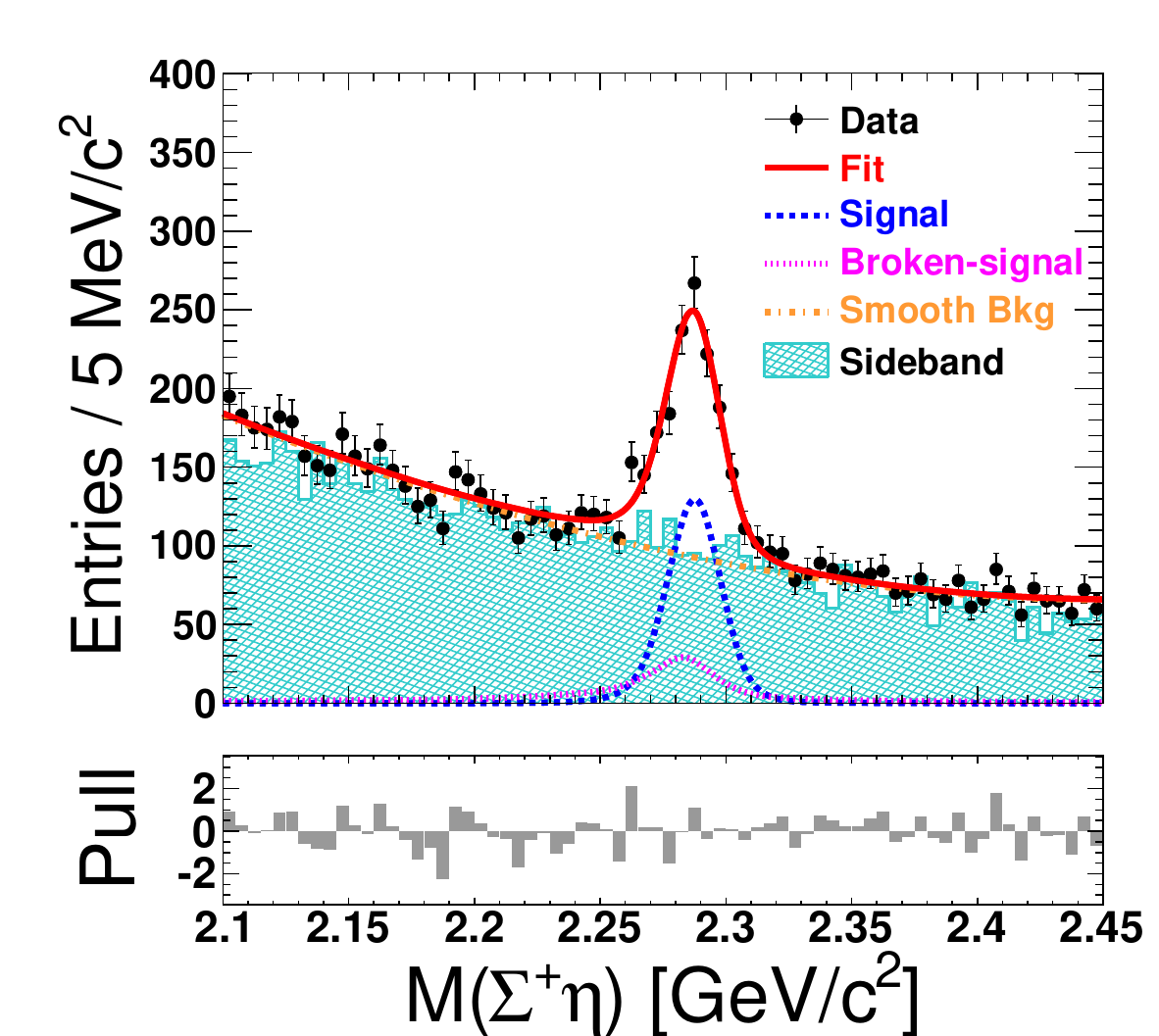}
    \put(21,65){\footnotesize{$\Lcp\to\Sigma^+\eta$}}
   \end{overpic}%
     \begin{overpic}[width=0.33\textwidth,height=0.25\textwidth]{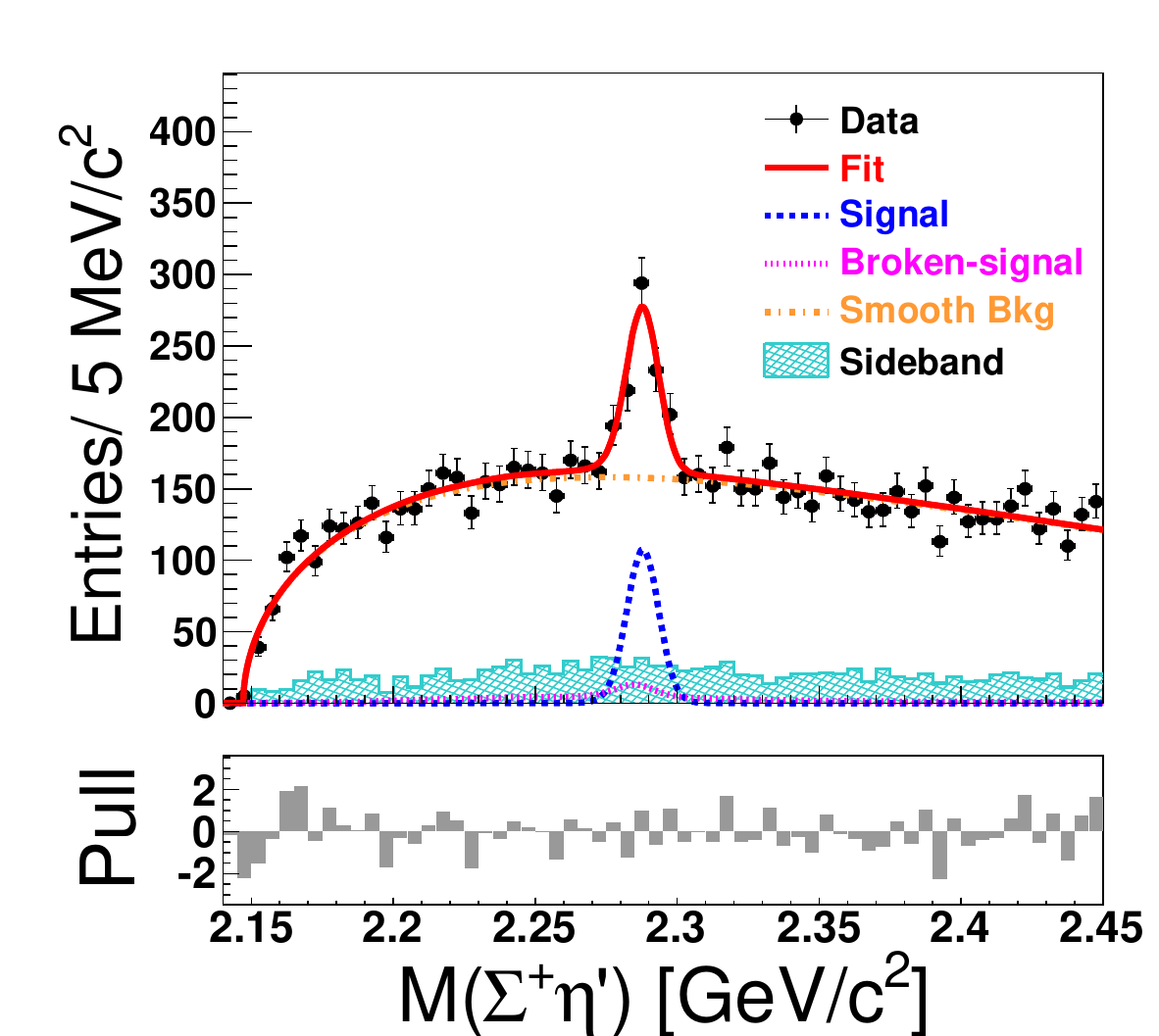}
    \put(21,65){\footnotesize{$\Lcp\to\Sigma^+\eta^{\prime}$}}
   \end{overpic}\\
   \vskip5pt
     \begin{overpic}[width=0.33\textwidth,height=0.25\textwidth]{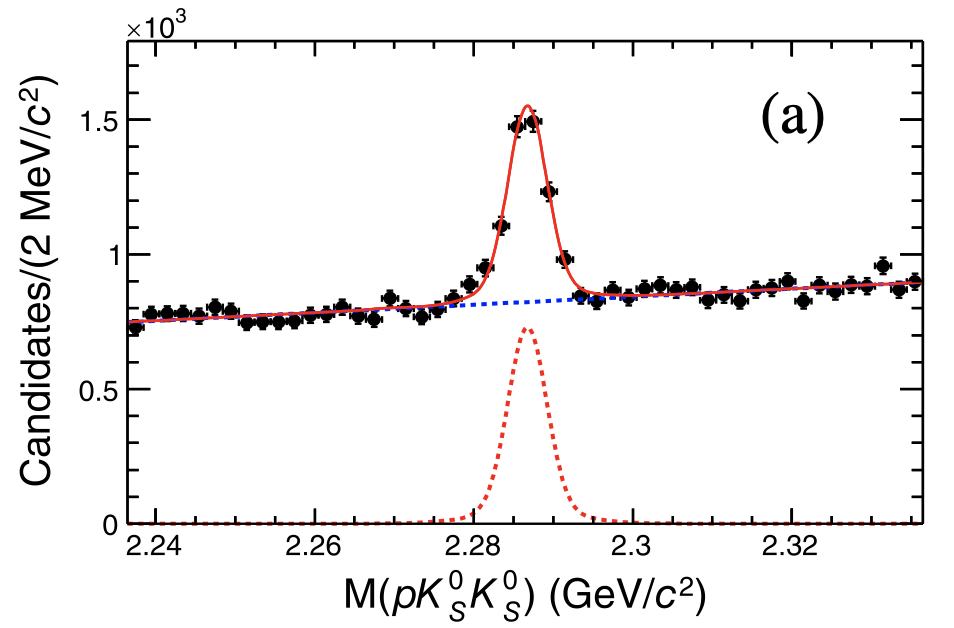}
    \put(18,65){\footnotesize{$\Lcp\to{}p\KS\KS$}}
   \end{overpic}%
    \begin{overpic}[width=0.33\textwidth,height=0.25\textwidth]{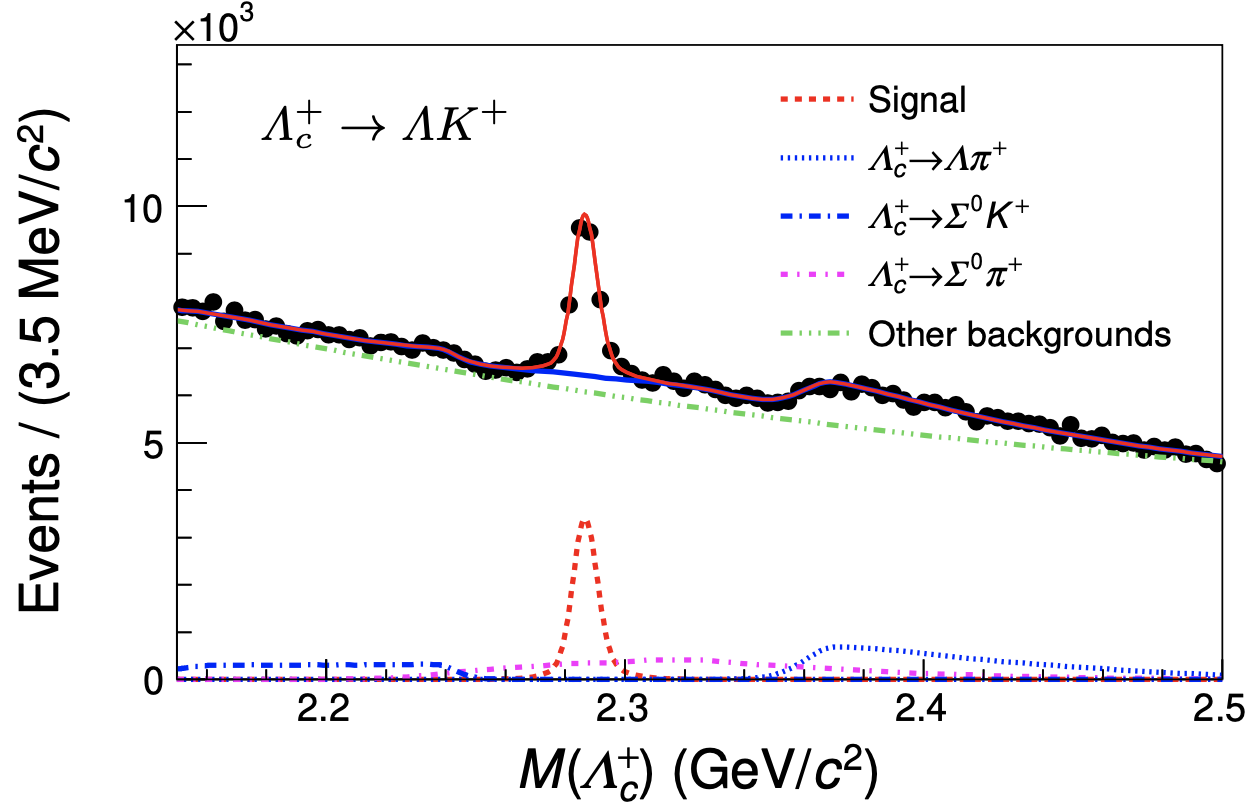}
    \put(20,30){\footnotesize{$\Lcp\to\Lambda\Kp$}}
    \end{overpic}%
     \begin{overpic}[width=0.33\textwidth,height=0.25\textwidth]{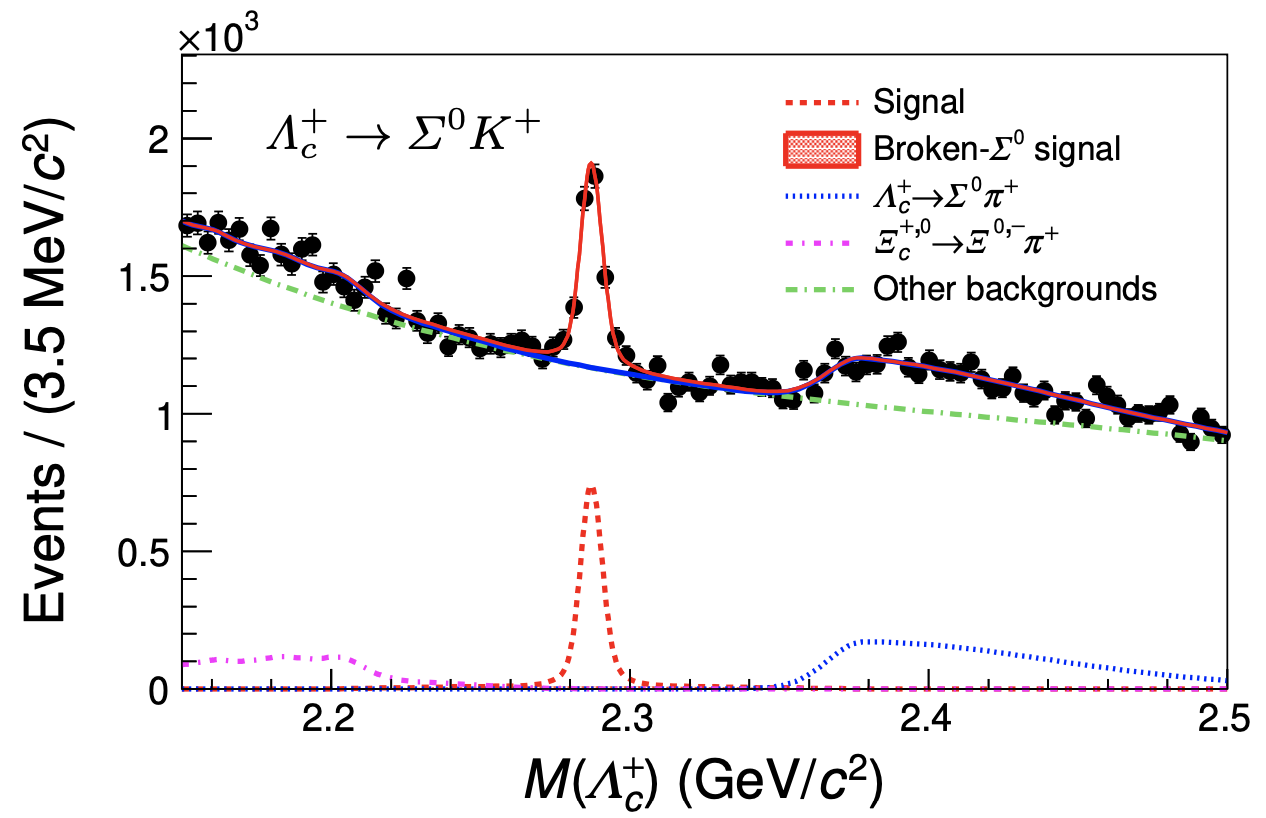}
    \put(20,30){\footnotesize{$\Lcp\to\Sigma^0\Kp$}}
    \end{overpic}%
   \vskip-5pt
\caption{\label{fig:BrLc}Invariant mass distributions of reconstructed $\Lcp$ candidates and their corresponding fit results for six decay modes at Belle.~\cite{Belle:2022pwd,Belle:2022bsi,Belle:2022uod,Belle:Xic0ToXi0Pi0}}
\end{center}
\end{figure}

The weak decays of charmed baryons provide an excellent platform for understanding QCD with transitions involving the charm quark. 
 The decay amplitudes consist of factorizable and non-factorizable contributions. 
 Experimentally, the study of charmed baryons is more challenging than that of charmed meson due to smaller experimental samples.
 Some CF decays are still poorly or note yet measured.
 Recently, Belle and Belle II reported many branching fractions of charmed baryons~\cite{Belle:2022pwd,Belle:2022bsi,Belle:2022uod,Belle:Xic0ToXi0Pi0}. 
 The distributions of invariant mass of reconstructed $\Lcp$ in six decay channels, and their corresponding fit results, are shown in Figure~\ref{fig:BrLc}.

We report the branching fractions of three CF and three SCS decays: 
\begin{eqnarray}
 \BR(\Lcp\to{}p\KS\eta)	& = & (4.35\pm 0.10\pm 0.20\pm 0.22)\times10^{-3}\,, \\ 
 \BR(\Lcp\to\Sigma^+\eta)	& = & (3.14\pm 0.35\pm 0.17\pm 0.25)\times10^{-3}\,,  \\
 \BR(\Lcp\to\Sigma^+\eta^{\prime})	& = & (4.16\pm 0.75\pm 0.25\pm 0.33)\times10^{-3}\,, \\
 \BR(\Lcp\to{}p\KS\KS)	& = & (2.35\pm 0.12\pm 0.07\pm 0.12)\times10^{-4}\,, \\ 
 \BR(\Lcp\to\Lambda\Kp)	& = & (6.57\pm 0.17\pm 0.11\pm 0.35)\times10^{-4}\,, \\
 \BR(\Lcp\to\Lambda\Kp) 	& = & (3.58\pm 0.19\pm 0.06\pm 0.19)\times10^{-4}\,,  
\end{eqnarray}
and five results for the $\Xi_c^0$ and $\Omega_c^0$ decays: 
\begin{eqnarray}
 \BR(\Xicz\to\Xi^0\piz)	& = & (6.9\pm 0.3\pm 0.5\pm 1.5)\times10^{-3}\,, \\
 \BR(\Xicz\to\Xi^0\eta)	& = & (1.6\pm 0.2\pm 0.2\pm 0.4)\times10^{-3}\,, \\
 \BR(\Xicz\to\Xi^0\eta^{'})	& = & (1.2\pm 0.3\pm 0.1\pm 0.3)\times10^{-3}\,, \\
 \frac{\BR(\Omega_c^0\to\Xi^-\pip)}{\BR(\Omega_c^0\to\Omega^-\pip)} & = & 0.253\pm 0.052\pm 0.030\,, \\ 
 \frac{\BR(\Omega_c^0\to\Xi^-\Kp)}{\BR(\Omega_c^0\to\Omega^-\pip)} & < & 0.070\,. 
\end{eqnarray}
All of these results are the first or most precise measurements to date.

\subsubsection{Decay asymmetry parameters of two-body decays of charmed baryons}
The decay asymmetry parameter $\alpha$ was introduced by Lee and Yang to study the parity-violating and parity-conserving amplitudes in weak hyperon decays.
In $1/2^+\to1/2^++0^-$, $\alpha \!\equiv\! {2\cdot {\rm Re}(S^{*}P)/(|S|^2 + |P|^2)}$, where $S$ and $P$ denote the parity-violating $S$-wave and parity-conserving $P$-wave amplitudes, respectively. 
Taking $\LcToLamHp,\Sigma^+h^0$ decays for example, the differential decay rate has a dependence on $\alpha$:
\begin{eqnarray}
\frac{dN(\LcToLamHp)}{d\cos\theta_{\Lambda}} \propto 1 + \alpha_{\Lcp}\alpha_{-}\cos\theta_{\Lambda}\,,
\end{eqnarray}
where $\alpha_{-}$ is hyperon decay asymmetry parameter.
For $\LcToSigHp$ decays, considering $\alpha(\Sigma^0\to\gamma\Lambda)$ is zero due to parity conservation for an electromagnetic decay, the differential decay rate is
\begin{eqnarray}
\frac{dN(\LcToSigHp)}{d\cos\theta_{\Sigma^0}d\cos\theta_{\Lambda}} \propto 1 - \alpha_{\Lcp}\alpha_{-}\cos\theta_{\Sigma^{0}}\cos\theta_{\Lambda}
\end{eqnarray}
By studying the hyperon helicity angle, we can extract $\alpha$ from charmed baryon decays. The results are listed in Tab.~\ref{tab:alpha}. 

 \begin{table}[!htbp] 
   \begin{center}   
     \tbl{\label{tab:alpha}Recent measurements of $\alpha$ at Belle~\cite{Belle:2022uod,Belle:2022bsi,Belle:2021crz,Belle:Xic0ToXi0Pi0,Belle:2021zsy}, with BESIII~\cite{BESIII:2019odb,BESIII:2023wrw,BESIII:2022udq}, CLEO~\cite{CLEO:2000lsg} and world average (W.A.)~\cite{bib:PDG2022} values.}
     {
  \begin{tabular}{lcc} \toprule
	Decay	                        		& Belle    					& Other experiments      \\ \midrule
	$\Lcp\to p \KS$                 		& --                    						& $0.18\pm0.45$\,~\cite{BESIII:2019odb}                 \\ 
	$\Lcp\to \Lambda K^{+}$         	& $-0.585\pm0.052$\,~\cite{Belle:2022uod}      	& --                   \\ 
	$\Lcp\to \Sigma^{0} K^{+}$	& $-0.54\,\,\,\pm0.20\,\,\,$\,~\cite{Belle:2022uod}       	& --                   \\   
	$\Lcp\to \Lambda \pi^{+}$	    	& $-0.755\pm0.006$\,~\cite{Belle:2022uod}     	& $-0.84\pm0.09$\,~\cite{bib:PDG2022}     \\
	$\Lcp\to \Sigma^{0}\pi^{+}$     	& $-0.463\pm0.018$\,~\cite{Belle:2022uod}     	& $-0.73\pm0.18$\,~\cite{BESIII:2019odb}     \\
	$\Lcp\to \Sigma^{+} \pi^{0}$    	& $-0.480\pm0.028$\,~\cite{Belle:2022bsi}    	& $-0.55\pm0.11$\,~\cite{bib:PDG2022}     \\
	$\Lcp\to \Sigma^{+} \eta  $     	& $-0.990\pm0.058$\,~\cite{Belle:2022bsi}     	& --                   \\				
	$\Lcp\to \Sigma^{+} \eta^{\prime}$  & $-0.460\pm0.067$\,~\cite{Belle:2022bsi} 	& --             \\
	$\Lcp\to \Xi^0\Kp$			& --  									& $+0.01\pm 0.16$\,~\cite{BESIII:2023wrw}	 	\\
	$\Lcp\to \Lambda \rho^+$		& --  									& $-0.76\pm 0.07$\,~\cite{BESIII:2022udq}   \\
	$\Lcp\to \Sigma^{\prime+}\piz$	& --  									& $-0.92\pm 0.09$\,~\cite{BESIII:2022udq}	            \\
	$\Lcp\to \Sigma^{\prime0}\pip$	& --  									& $-0.79\pm 0.11$\,~\cite{BESIII:2022udq}          \\ \midrule
        $\Xi_{c}^{0}\to \Xi^{-} \pi^{+}$    	     & $-0.63\pm0.03$\,~\cite{Belle:2021crz} 	& $-0.56\pm0.40$\,~\cite{CLEO:2000lsg} \\ 
        $\Xi_{c}^{0}\to \Xi^{0} \pi^{0}$    	     & $-0.90\pm0.27$\,~\cite{Belle:Xic0ToXi0Pi0} 	& -- \\ 
        $\Xi_{c}^{0}\to \Lambda \Kstarzb$  & 	$+0.15\pm0.22$\,~\cite{Belle:2021zsy}   	&  -- \\
        $\Xi_{c}^{0}\to \Sigma^{+}K^{*-}$   &  $-0.52\pm0.30$\,~\cite{Belle:2021zsy}    	&  -- \\	\bottomrule
  \end{tabular}
}
  \end{center}
\end{table}

\section{Search for rare or forbidden decays in charm sector}
\begin{figure}[!htpb]
\begin{center}
     \begin{overpic}[width=0.33\textwidth,height=0.26\textwidth]{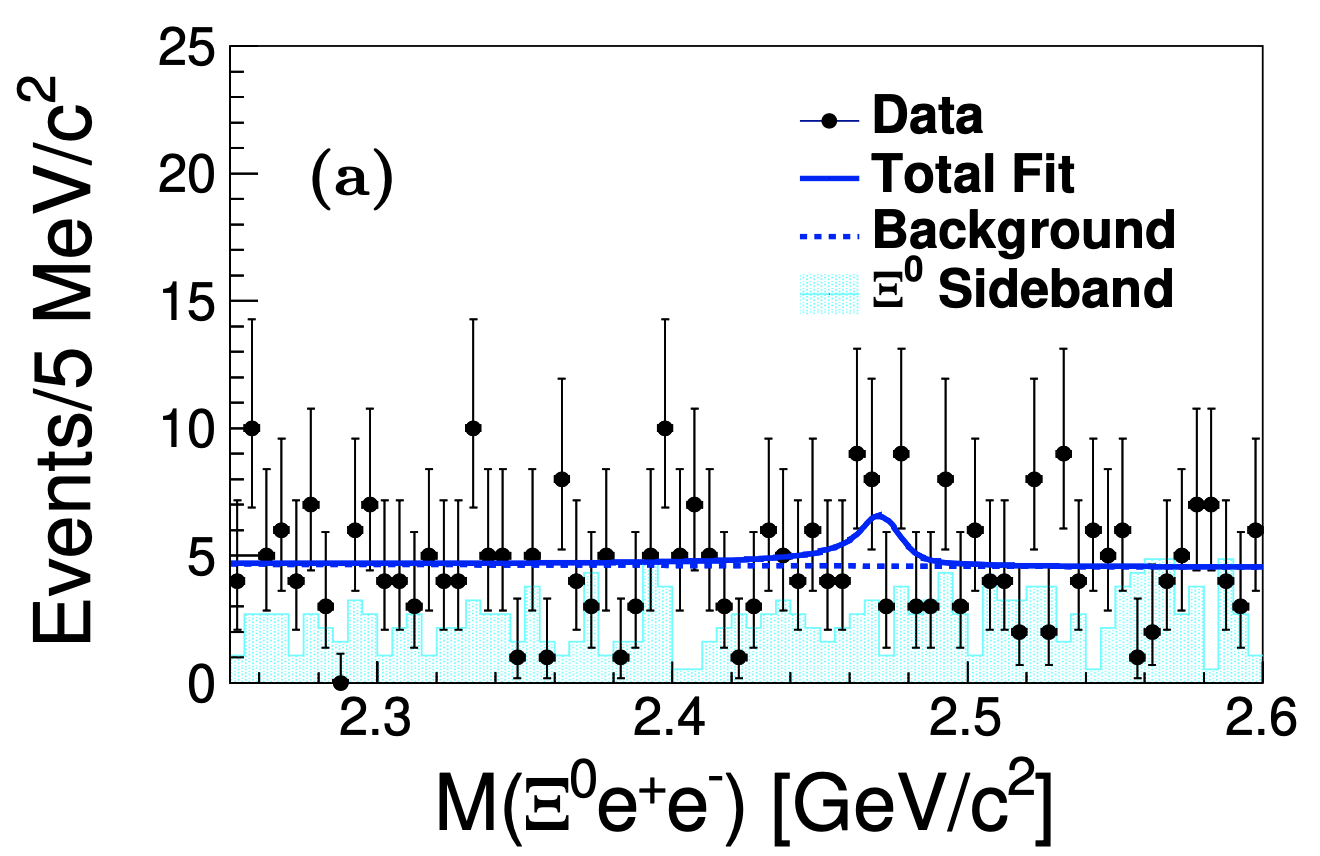}
    \end{overpic}%
     \begin{overpic}[width=0.33\textwidth,height=0.26\textwidth]{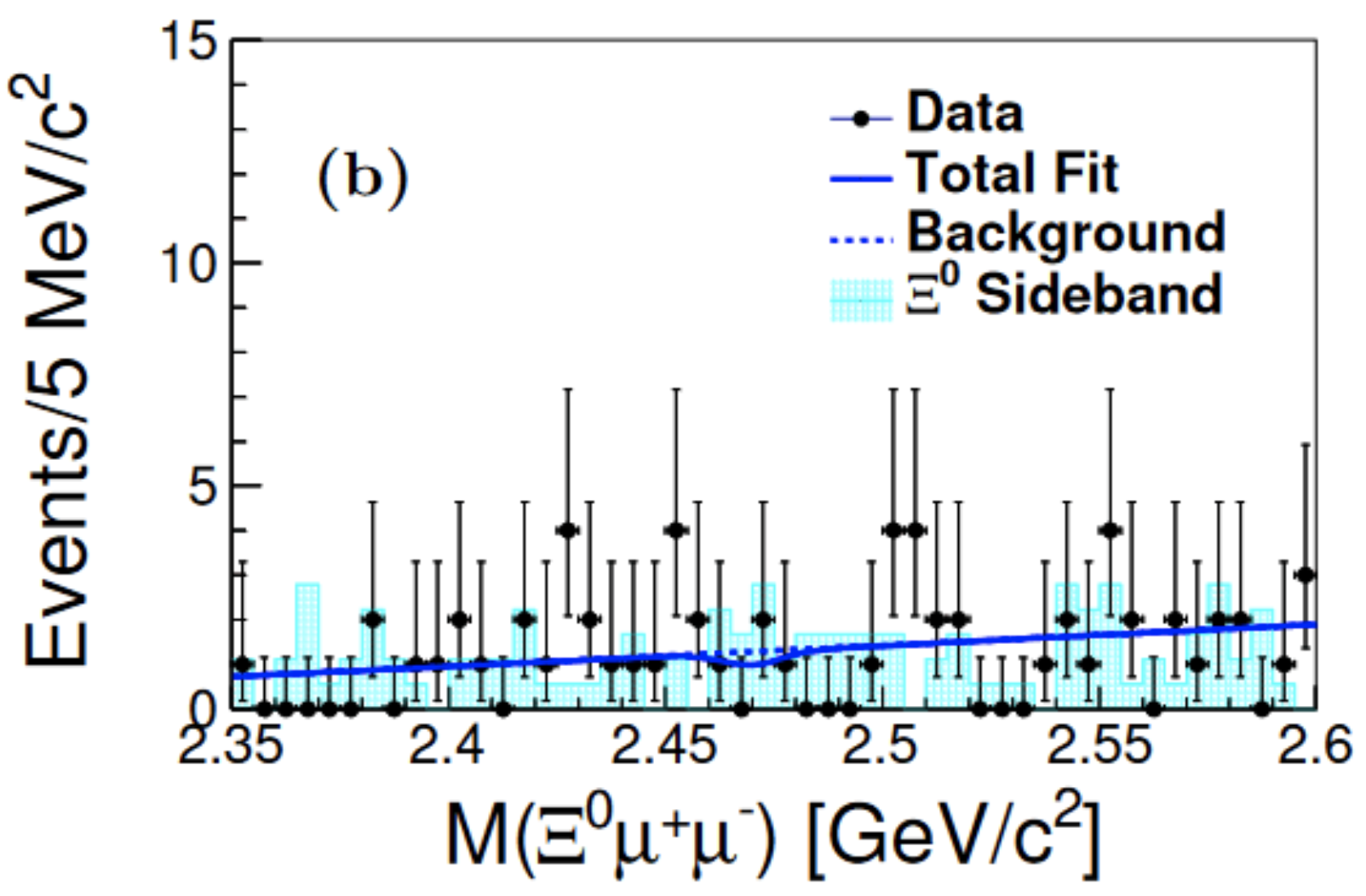}
    \end{overpic}%
     \begin{overpic}[width=0.33\textwidth,height=0.26\textwidth]{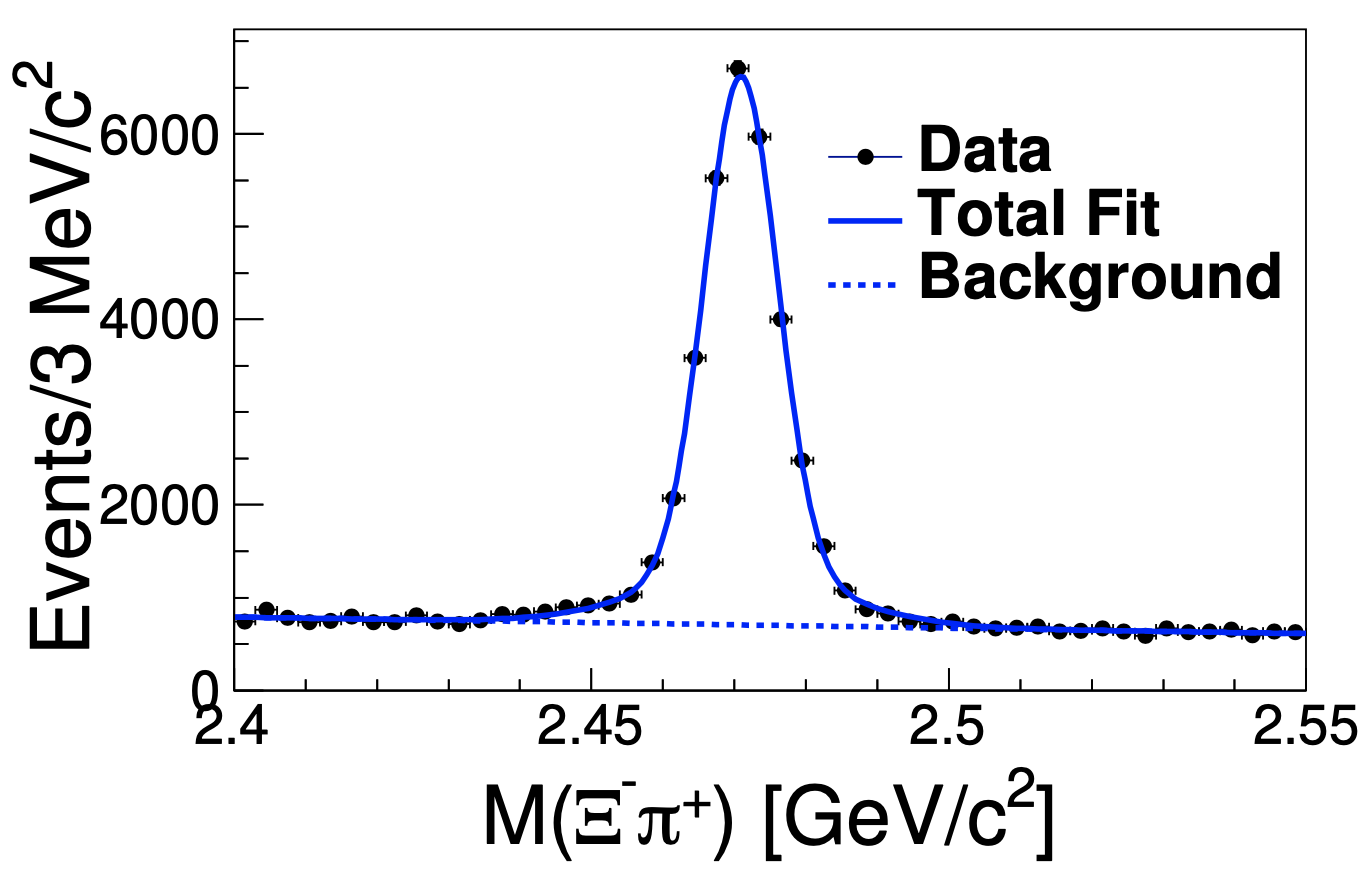}
    \end{overpic}%
    \vskip-5pt
\caption{\label{fig:Xic0Rare}The invariant mass of reconstructed $\Xicz$ candidates for signal modes $\Xi_c^0\to\Xi^0\ell^+\ell^-$ and reference mode $\Xicz\to\Xim\pip$ at Belle.~\cite{Belle:2023ngs}}
\end{center}
\end{figure}

In the Standard Model (SM), the weak-current interaction has an identical coupling to all lepton generations (Lepton Flavor Universality (LFU)). 
LFU can be tested in semi-leptonic decays, such as $\Xi_c^0\to\Xi^0\ell^+\ell^-$ where a comparison of $\ell=e$ and $\mu$ decay rates would comprise such a test.
Recently Belle reported a search for $\Xi_c^0\to\Xi^0\ell^+\ell^-$ based on the Belle full data set~\cite{Belle:2023ngs}.
The fits of invariant mass of reconstructed $\Xicz$ candidates for signal modes and reference mode are shown in Figure~\ref{fig:Xic0Rare}. 
The upper limits on branching fractions relative to reference mode $\Xicz\to\Xim\pip$ are measured to be 
$\frac{\BR(\Xi_c^0\to\Xi^-e^+e^-)}{\BR(\Xi_c^0\to\Xi^-\pip)}<6.7\times 10^{-3}$
and $\frac{\BR(\Xi_c^0\to\Xi^-\mu^+\mu^-)}{\BR(\Xi_c^0\to\Xi^-\pip)}<4.3\times 10^{-3}$.
A more precise analysis based on larger data samples collected by Belle~II is expected in the future.

Baryon number violation (BNV) is one of the crucial conditions to create matter-antimatter asymmetry as observed in the universe. 
Several grand unified theories, supersymmetry and other SM extensions propose BNV processes of nucleons.
The $D\to p\ell$ decays violate baryon (B) and lepton (L) numbers but their difference is conserved ($\Delta(B-L)=0$). 
The previous stringent limit is $\BR(D^0\to\bar{p}e^+)<1.2\times 10^{-6}$ at a 90\% C.L. and recent BESIII result is $\BR(D^0\to pe^-)<2.2\times 10^{-6}$.
Recently, Belle reported a stricter upper limits: $(5-8)\times 10^{-7}$ dependent on the decay modes, as shown in Table~\ref{tab:DTopl}.
 \begin{table}[!htbp] 
   \begin{center}   
   \tbl{\label{tab:DTopl}Reconstruction efficiency ($\eff$), signal yield ($N_S$), signal significance ($S$), upper limit on the signal yield ($N_{p\ell}^{\rm UL}$), and branching fraction ($\BR$) at 90\% confidence level for baryon number violating decay modes.}{
  \begin{tabular}{cccccc} \toprule
  Decay mode~~ 	& ~$\eff$(\%)~ & ~$N_{S}$~		& ~$S(\sigma)$~	& $N_{p\ell}^{\rm UL}$~ & ~$\BR$ ($10^{-7})$ \\ \hline 
  $D^0\to pe^-$	  	&  10.2		 & $-6.4\pm 8.5$	& --				& 17.5		& $<5.5$		\\
  $\bar{D}^0\to pe^-$	&  10.2		 & $-18.4\pm 23.0$	& --				& 22.0		& $<6.9$		\\
  $D^0\to\bar{p}e^+$	&  9.7		 & $-4.7\pm 23.0$	& --				& 22.0		& $<7.2$		\\
  $\bar{D}^0\to\bar{p}e^+$	&  9.6	 & $7.1\pm 9.0$	& 0.6				& 23.0		& $<7.6$		\\
  $D^0\to p\mu^-$	  	&  10.7		 & $11.0\pm 23.0$	& 0.9			& 17.1		& $<5.1$		\\
  $\bar{D}^0\to p\mu^-$	&  10.7		 & $-10.8\pm 27.0$	& --			& 21.8		& $<6.5$		\\
  $D^0\to\bar{p}\mu^+$	&  10.5		 & $-4.5\pm 14.0$	& --			& 21.1		& $<6.3$		\\
  $\bar{D}^0\to\bar{p}\mu^+$	&  10.4	 & $16.7\pm 8.8$	& 1.6			& 21.4		& $<6.5$		\\ \botrule
  \end{tabular}
}
  \end{center}
\end{table}

\section{Charm $\CP$ violation searches}
The violation of $\CP$-symmetry, the combination of charge conjugation symmetry and parity asymmetry, is essential for elucidating the matter-antimatter asymmetry in the universe. 
In the Standard Model (SM) of particle physics, the sole source of $\CP$ violation~(CPV) arises from a single complex phase in the Cabibbo-Kobayashi-Maskawa matrix. 
However, this source is insufficient to account for the observed matter-antimatter asymmetry. 
Therefore, we need new CPV sources beyond the SM. 
Charm CPV in the SM is very small, at level of $\mathcal{O}(10^{-3})$ or smaller, but new physics (NP) may enhance it. Therefore, a study of charm CPV may help to test the SM and act as a sensitive probe for NP.
Experimentally, we have only one CPV observation in charm sector: $\Delta \Acp(D^0\to\Kp\Km,\pip\pim)=(-15.4\pm2.9)\times 10^{-4}$ ($5.3\sigma$) from LHCb. To understand such CPV, we need to work on more channels and improve the precision of measured $\CP$ asymmetries. 
On the other hand, CPV has been observed in the open-flavored meson sector, but not yet in the baryon sector. 
Baryogenesis, the process by which the baryon-antibaryon asymmetry of the universe developed, is directly related to baryon CPV. 
Discovering the CPV in charmed baryon decays is correctly one of the main targets of charm physics. 
Recently we have reported CPV searches in four-body decays of charmed mesons, and $\alpha$-induced CPV and direct CPV in $\Lcp$ two-body decays.

\subsection{CPV in four-body decays of charmed mesons}
Sensitivity to CPV varies with the decay channel, motivating CPV searches in diverse charm decays. 
The $D$ four-body decays, with large branching fractions and involving various intermediate processes, provide a good platform for CPV searches. 
CPV in $D$ four-body decay was probed with triple-product asymmetries by the FOCUS, BABAR, LHCb and Belle experiments. 
The triple-product~(TP) is defined in the $D$ rest frame using the momenta of three particles in the final state, ${C_{\rm TP} = \vec{p}_i\cdot (\vec{p}_j \times \vec{p}_k)}$ for $D\to{}P_iP_jP_kP_l$ decays, 
and satisfies $\CP(C_{\rm TP})=-\overline{C}_{\rm TP}$. 
The sign of $C_{\rm TP}$ denotes whether the $\vec{p}_i$ points ``upward'' or ``downward'' in the plane defined by $\vec{p}_j$ and $\vec{p}_k$, 
therefore, its asymmetry is called an up-down asymmetry. 
The TP asymmetries in $D^+$ and $D^-$ decays are defined as 
\begin{eqnarray}
A_{T}(D^+) & = & \dfrac{N_+(C_T>0) - N_+(C_T<0)}{N_+(C_T>0) + N_+(C_T<0)}\,, \\
\overline{A}_{T}(D^-) & = & \dfrac{N_-(-\overline{C}_T>0) - N_-(-\overline{C}_T<0)}{N_-(-\overline{C}_T>0) + N_-(-\overline{C}_T<0)} \,.
\end{eqnarray}
And their difference is assigned as a $\CP$-violating parameter, i.e. $a_{\CP}^{\rm T-odd} = {\frac{1}{2}\cdot (A_{T}(D^+) - \overline{A}_{T}(D^-))}$.
This parameter $a_{\CP}^{\rm T-odd}\propto\sin\phi\cos\delta$, where $\phi$ and $\delta$ are the weak and strong phase differences, respectively, between at least two amplitudes contributing to the decay. 
The $a_{\CP}^{\rm T-odd}$ has its largest value when $\delta=0$, while a non-zero direct $\CP$ asymmetry requires $\delta\ne0$, therefore  $a_{\CP}^{\rm T-odd}$ is an observable complementary to direct $\CP$ asymmetry. 

\begin{figure}[!htpb]
\begin{center}
  \begin{overpic}[width=0.8\textwidth]{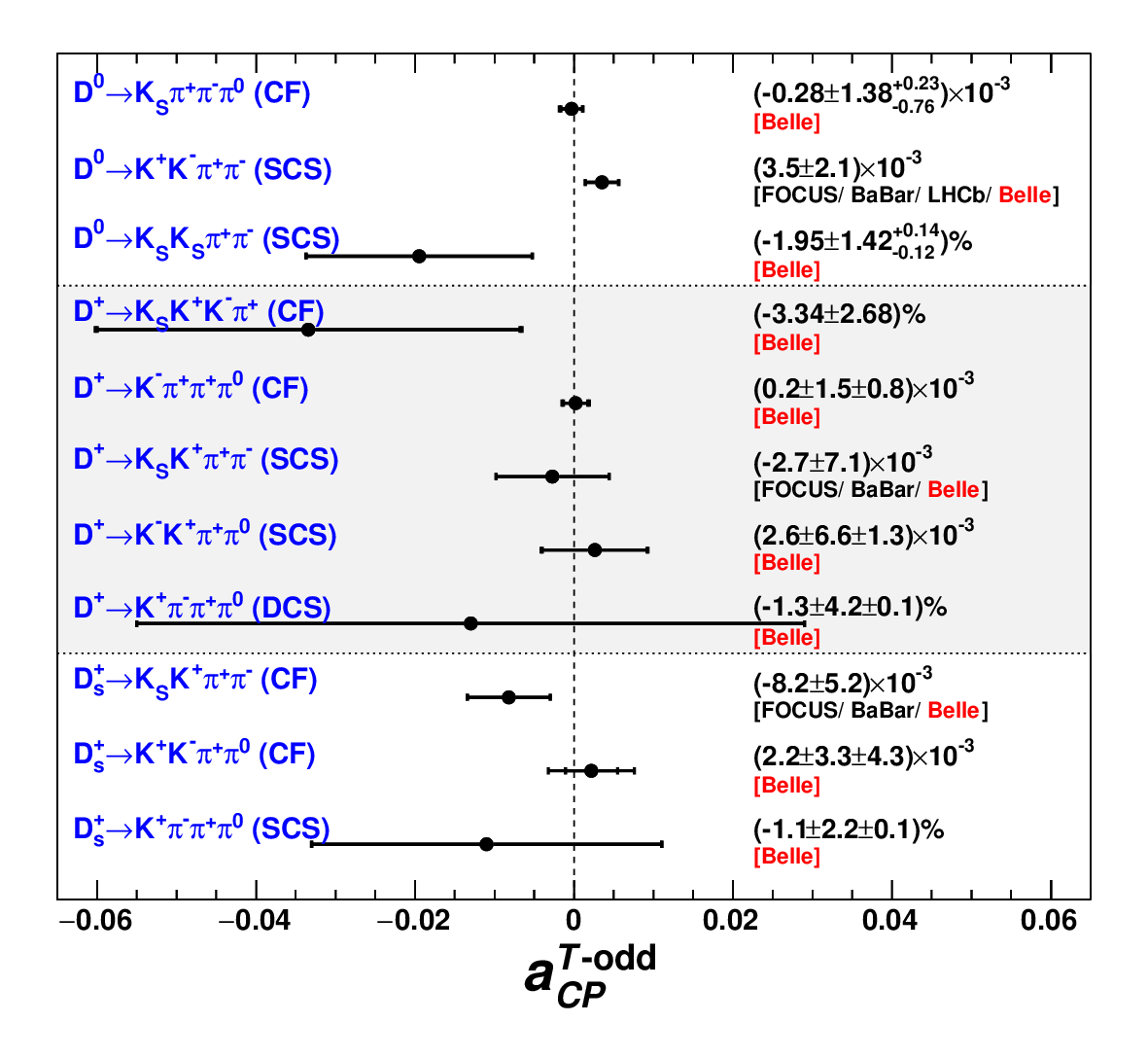}%
  \end{overpic}  
  \vskip-10pt
  \caption{\label{fig:Todd}Belle results for $a_{\CP}^{\rm T-odd}$ along with other measurements for $D^0$ and $D_{(s)}^+$ decays. For decays in which more than one measurement has been made, the world average value is plotted.}
\end{center}
\end{figure}

Recently Belle searched for CPV with TP asymmetries in the decays of $D^0\to\KS\KS\pip\pim$~\cite{Belle:2022xof}, $D_{(s)}^+\to\KS h^+\pip\pim$~\cite{Belle:2023bzn}, and $D_{(s)}^+\to Kh\pip\piz$~\cite{Belle:2023str}. They are listed in Figure~\ref{fig:Todd}.
Most of these $\ATodd$ results from Belle are first or most precise measurements. 

\subsection{CPV in $\Lcp\to\Lambda K^+,\Sigma^0 K^+$} 
Recently, a search for direct CPV and $\alpha$-induced CPV in $\Lcp\to\Lambda K^+,\Sigma^0 K^+$ was reported based on the Belle full data set~\cite{Belle:2022uod}.

For SCS decay, for example $\Lcp\to\Lambda\Kp$, the raw asymmetry includes several sources: 
\begin{eqnarray}
A_{\rm raw} = A_{\CP}^{\Lcp\to\Lambda\Kp} + A_{\CP}^{\Lambda\to{}p\pim} + A_{\eff}^{\Lambda} + A_{\eff}^{\Kp} + A_{\rm FB}^{\Lcp} 
\end{eqnarray}
where $\Acp^{\LcToLamKp}$ ($\Acp^{\Lambda\to{}p\pim}$) is the $\CP$ asymmetry associated with $\Lcp$ ($\Lambda$) decay; 
$A_{\varepsilon}^{\Lambda}$ is an asymmetry arising from detection efficiencies of $\Lambda$ and $\Lbar$; 
$A_{\varepsilon}^{\Kp}$ is the $\Kp$ reconstruction and identification asymmetry and can be removed by weighting $w_{\Lcp,\Lcm} = 1 \mp A_{\varepsilon}^{\Kp}[\cos\theta,\,p_T]$;
$A_{\rm FB}^{\Lcp}$ arises from the forward-backward asymmetry of $\Lcp$ production due to $\gamma$-$Z^0$ interference and higher-order QED effects in ${\epem\to c\cbar}$ collisions.
We use the corresponding CF modes, $\Lcp\to\Lambda\pip$ and $\Lcp\to\Sigma^0\pip$, as reference modes to remove the common asymmetry sources: $A_{\CP}^{\Lambda\to{}p\pim}$, $A_{\eff}^{\Lambda}$ and $A_{\rm FB}^{\Lcp}$. Under the current precision, the CPV in charm CF mode is consistent with zero, i.e. $A_{\CP}^{\Lcp\to\Lambda\pip}=0$.
Finally, we have first results of a search for direct $\CP$ asymmetry in two-body SCS decays of charmed baryons:
\begin{eqnarray}
\Acp^{\rm dir}(\LcToLamKp) & = & (+2.1 \pm 2.6 \pm 0.1)\%\,, \\
\Acp^{\rm dir}(\LcToSigKp) & = & (+2.5 \pm 5.4 \pm 0.4)\%\,.
\end{eqnarray}

For $\LcToLamHp$ decays, the differential decay rate depends on $\alpha$ parameters and one helicity angle:
\begin{eqnarray}
\frac{dN}{d\cos\theta_{\Lambda}} \propto 1+\alpha_{\Lcp}\alpha_{-}\cos\theta_{\Lambda}\,, \label{eqn:alpha_LcToLamHp}
\end{eqnarray}
where $\alpha_{\Lcp}$ is the decay asymmetry parameter of $\LcToLamHp$, and $\theta_{\Lambda}$ is the angle between the proton momentum and the direction opposite the $\Lcp$ momentum in the $\Lambda$ rest frame.

\begin{figure}[!htpb]
\begin{center}
  \begin{overpic}[width=0.45\textwidth]{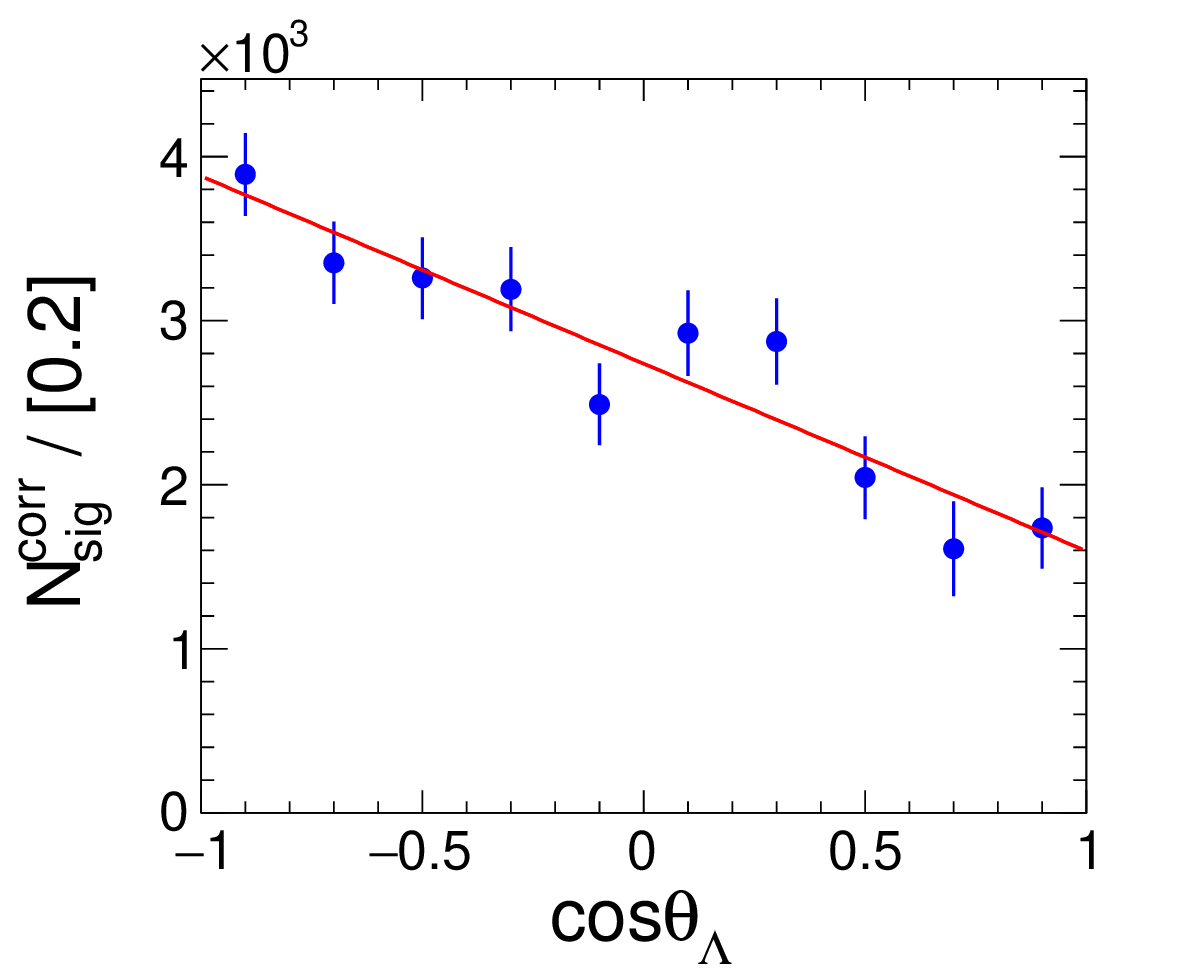}%
  \put(50,65){\small{$\LcToLamKp$}}%
  \end{overpic}%
  \begin{overpic}[width=0.45\textwidth]{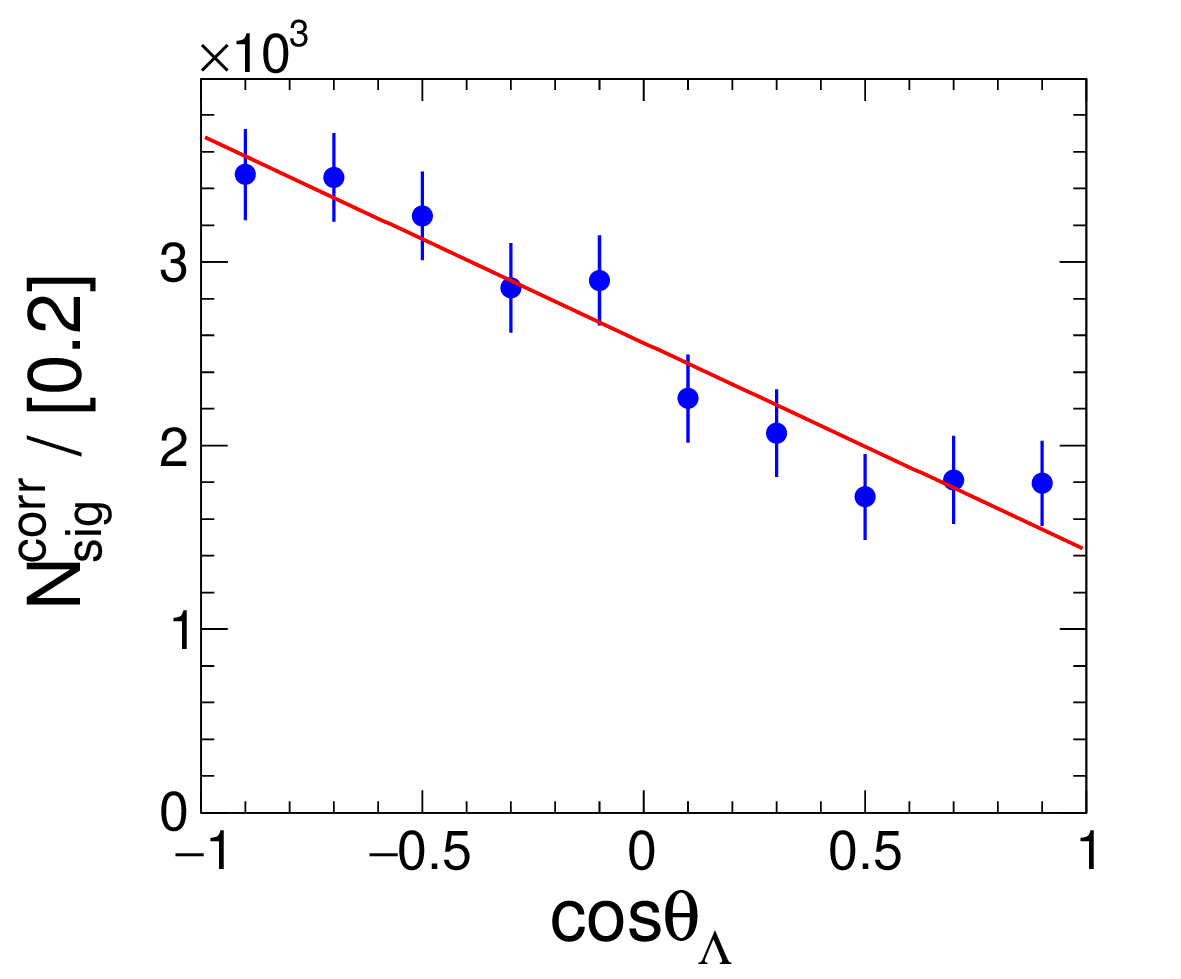}%
  \put(50,65){\small{$\Lcm\to\overline{\Lambda}\Km$}}%
  \end{overpic}
    \vskip-5pt
\caption{\label{fig:alpha_LcToLamHp}The $\cos\theta_{\Lambda}$ distribution of $\Lcp\to\Lambda\Kp$ after efficiency-correction. We fit with a linear function of $1+\alpha_{\Lambda_c^{\pm}}\alpha_{\mp}\cos\theta_{\Lambda}$ with goodness-of-fit $\chi^2/9=1.04,\,0.57$, respectively, at Belle.~\cite{Belle:2022uod}}
\end{center}
\end{figure}

For $\LcToSigHp$ decays, considering $\alpha(\Sigma^0\to\gamma\Lambda)$ is zero due to parity conservation for an electromagnetic decay, the differential decay rate is given by
\begin{eqnarray}
\frac{dN}{d\cos\theta_{\Sigma^0}d\cos\theta_{\Lambda}} \propto 1 - \alpha_{\Lcp}\alpha_{-}\cos\theta_{\Sigma^{0}}\cos\theta_{\Lambda}\,, \label{eqn:alpha_LcToSigHp}
\end{eqnarray}
where $\theta_{\Lambda}$ ($\theta_{\Sigma^0}$) is the angle between the proton ($\Lambda$) momentum and the direction opposite the $\Sigma^0$ ($\Lcp$) momentum in the $\Lambda$ ($\Sigma^0$) rest frame.
Since $\alpha$ is a CP-odd observable, the corresponding $\CP$-violating parameter is defined as 
\begin{eqnarray}
A_{\CP}^{\alpha} = \frac{ \alpha_{\Lcp} + \alpha_{\Lcm} }{\alpha_{\Lcp} - \alpha_{\Lcm} }\,.
\end{eqnarray}
Under $\CP$ conservation, we have $\alpha_{\Lcp} = - \alpha_{\Lcm}$. 
We measured the $\alpha$-parameters for the separate $\Lcp$ and $\Lcm$ samples, as shown in Figure~\ref{fig:alpha_LcToLamHp} for $\Lcp\to\Lambda\Kp$,
 and calculate the $\alpha$-induced CPV parameter $A_{\CP}^{\alpha}$. 
We have 
\begin{eqnarray}
\Acp^{\alpha}(\LcToLamKp) & = & -0.023\pm 0.086\pm 0.071 \,, \\
\Acp^{\alpha}(\LcToSigKp) & = & +0.08\pm 0.35\pm 0.14 \,.
\end{eqnarray}
No evidence of CPV is found in these two decays.

We also probe the $\Lambda$-hyperon CPV in CF decays $\LcToLamPip$ and $\LcToSigPip$, inspired by a theoretical paper~\cite{Wang:2022tcm}. 
The $\Lambda$-hyperon $\CP$ asymmetry $\Acp^{\alpha}({\Lambda\to{}p\pim})$ can be extracted from the total $\alpha$-induced $\CP$ asymmetry of $\Lcp$ decay chain: 
\begin{eqnarray}
\Acp^{\alpha}({\rm total}) \equiv \frac{\alpha_{\Lcp}\alpha_{-} - \alpha_{\Lcm}\alpha_{+}}{\alpha_{\Lcp}\alpha_{-} +\alpha_{\Lcm} \alpha_{+}} = A_{\CP}^{\alpha} (\Lambda \to p \pim)\,.
\end{eqnarray}
for Cabibbo-favored (CF) decays $\Lcp\to(\Lambda,\,\Sigma^0)\pip$, $\alpha_{\Lcp}\!=\!-\alpha_{\Lcm}$ since no $\CP$ asymmetry is expected in the SM.
CPV in hyperon decays is predicted to be at the level of $\mathcal{O}(10^{-4})$ or smaller in the SM~\cite{Donoghue:1985ww,Donoghue:1986hh,Tandean:2002vy,Salone:2022lpt} 
and can be enhanced to reach the level of $10^{-3}$ in some new physics models~\cite{Chang:1994wk,He:1999bv,Chen:2001cv,Tandean:2003fr,Salone:2022lpt}.
The average value of $\Acp^{\alpha}({\Lambda\to{}p\pim})$ in two such CF modes is calculated to be 
\begin{eqnarray}
A_{\CP}^{\alpha} (\Lambda \to p \pim) = +0.013 \pm 0.007 \pm 0.011\,.
\end{eqnarray}
This is the first measurement of hyperon CPV searches in CF charm decays.
No evidence of $\Lambda$-hyperon CPV is found.

\section{Summary}
\begin{figure}[!htpb]
\begin{center}
  \begin{overpic}[width=0.8\textwidth]{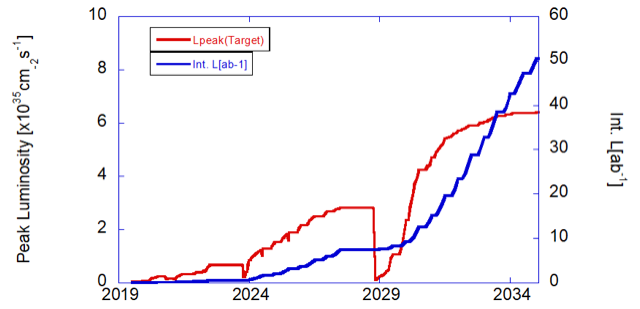}%
  \end{overpic}%
    \vskip-5pt
  \caption{\label{fig:B2lumin}Luminosity projection with plans up to spring 2034 at SuperKEKB. }
\end{center}
\end{figure}

Belle continues to produce the fruitful charm results, even though its data taking finished $13$ years ago. 
Belle II has joined the game since 2019. Now a dataset with 427~$\invfb$ is available. 
We reported some recent results on measurements of $\BR$ and $\alpha$, 
CPV searches in the charmed meson and baryon decays, and several searches for rare or forbidden decays.
By utilizing the early dataset at Belle~II, we obtain the world's best $\tau(D^{0,+})$, $\tau(D_s^+)$, and $\tau(\Lcp)$, and confirmation of the LHCb $\tau(\Omega_c^0)$ result.
More charm results based on a combined dataset of 1.4~$\invab$ at Belle and Belle~II will be forthcoming. 
The scheduled luminosity accumulations, as shown in Figure~\ref{fig:B2lumin}, promise the fruitful charm results at Belle~II in the future. 

\appendix



\end{document}